\documentclass[trackchanges,twocolumn]{aastex701}

\usepackage{soul}
\usepackage{amsmath}
\usepackage{tabularx}

\begin{document}

\title{Fast and furious -- High-velocity features of Ca and Si in the spectra of Type Ia supernovae \footnote{Based on observations obtained with the Hobby-Eberly Telescope (HET), which is a joint project of the University of Texas at Austin, the Pennsylvania State University, Ludwig-Maximillians-Universitaet Muenchen, and Georg-August Universitaet Goettingen. The HET is named in honor of its principal benefactors, William P. Hobby and Robert E. Eberly.}}

\correspondingauthor{Zs\'ofia Bora}
\email{bora.zsofia@csfk.hun-ren.hu}

\author[0000-0001-6232-9352]{Zs\'ofia Bora}
\affiliation{HUN-REN CSFK Konkoly Observatory, MTA Center of Excellence,
Konkoly Thege M. \'ut 15-17, Budapest, 1121, Hungary}
\affiliation{ELTE E\"otv\"os Lor\'and University, Institute of Physics and Astronomy, 
P\'azm\'any P\'eter s\'et\'any 1A, Budapest 1117, Hungary}
\email{bora.zsofia@csfk.hun-ren.hu}

\author[0000-0001-8764-7832]{J\'ozsef Vink\'o}
\affiliation{HUN-REN CSFK Konkoly Observatory, MTA Center of Excellence, 
Konkoly Thege M. \'ut 15-17, Budapest, 1121, Hungary}
\affiliation{Department of Experimental Physics, Institute of Physics, University of Szeged, D\'om t\'er 9, Szeged, 6720 Hungary}
\affiliation{ELTE E\"otv\"os Lor\'and University, Institute of Physics and Astronomy,  
P\'azm\'any P\'eter s\'et\'any 1A, Budapest 1117, Hungary}
\affiliation{Department of Astronomy, University of Texas at Austin,
2515 Speedway, Stop C1400, Austin, TX, 78712-1205, USA}
\email{vinko.jozsef@csfk.hun-ren.hu}

\author[0000-0003-1349-6538]{J. Craig Wheeler}
\affiliation{Department of Astronomy, University of Texas at Austin,
2515 Speedway, Stop C1400, Austin, TX, 78712-1205, USA}
\email{wheel@astro.as.utexas.edu}

\author[0000-0002-8770-6764]{R\'eka K\"onyves-T\'oth}
\affiliation{HUN-REN CSFK Konkoly Observatory, MTA Center of Excellence, Konkoly Thege M. \'ut 15-17, Budapest, 1121, Hungary}
\email{konyvestoth.reka@csfk.hun-ren.hu}

\author[0000-0002-2966-3508]{G. H. Marion}
\affiliation{Department of Astronomy, University of Texas at Austin,
2515 Speedway, Stop C1400, Austin, TX, 78712-1205, USA}
\email{ghmarion@gmail.com}

\begin{abstract}

High velocity absorption lines of Ca II and Si II (high-velocity features, HVFs) are common in the early spectra of Type Ia supernovae (SNe~Ia), but their physical origin is still poorly understood.
We used 181 optical spectra of 56 SNe~Ia that show HVFs in at least the Ca II NIR triplet to better understand the nature of HVFs.

To model the profiles of HVFs we applied Gaussian fitting as well as spectrum synthesis modeling with the SYNOW code. We found that Gaussian fitting usually underestimates the velocity of HVFs compared to SYNOW, and derived a linear formula for correction. In accord with many previous studies, the observed properties of HVFs (velocity, strength, duration) were found to correlate with the general observables of SNe Ia (decline rate, spectral subtype). Brighter, slow-declining SNe usually exhibit stronger, higher velocity HVFs, while fainter objects typically show weak, low velocity HVFs. SNe Ia in early-type galaxies showed weaker and slower HVFs compared to events in late-type galaxies in our sample. 

Comparing these observational findings with theoretical predictions, 
we conclude that the overall physical picture of HVFs is quite complex, and could be explained by multiple models. A double detonation explosion model is promising as it is inherently capable of explaining the creation of high-velocity shells that may produce HVFs, but an asymmetric delayed detonation model or interaction with circumstellar material cannot be ruled out.

\end{abstract}

\keywords{\uat{Type Ia supernovae}{1728}}

\section{Introduction}

Type Ia supernovae (SNe Ia) are believed to be thermonuclear explosions of carbon-oxygen white dwarfs (C/O WDs) in binary systems. The nature of the binary partner is not clear, and could be either a non-degenerate companion star (single degenerate, SD scenario), or another WD (double degenerate, DD scenario). In an SD scenario, the WD accretes material from its non-degenerate companion and approaches the Chandrasekhar mass ($M_{\rm Ch}$), where it eventually ignites a thermonuclear C/O explosion \citep{whelaniben73}. In this scenario, the WD can also gain angular momentum along with the material, resulting in a rapidly spinning WD that can  exceed $M_{\rm Ch}$ thanks to the extreme centrifugal forces. As such a WD loses its momentum over the years and the rotation slows, it eventually reaches critical density and explodes. These so called spin-up/spin-down models would result in near-, or even super-Chandra explosions \citep{distefano11}. In a DD configuration, two WDs merge or collide, which then leads to a similar thermonuclear explosion \citep{ibentutukov84, webbink84, pakmor12}. 

See e.g. \citet{ruiter25} for a recent, comprehensive review on SN Ia progenitors and explosion mechanisms.

Because of their high intrinsic brightness and substantially uniform spectra and light curves (LCs), SNe Ia have long been used as standardizable candles for cosmological distance measurements \citep{riess98, perlmutter99}. They can be used for this purpose because of the existence of an important connection between their peak brightness and the decline rate of their LCs, namely the Phillips relation \citep{phillips93}. The decline rate measures the rate at which the SN dims after maximum brightness, and is generally characterized by $\Delta m_{15}(\rm B)$, which gives the change in B magnitude between peak and 15 days after that. Other parameters that describe the shape of the LC are also in use, such as SALT $x_1$, a fitting parameter of the SALT LC model that measures the ``stretch" of the LC \citep{guy07, betoule14, kenworthy21}, or $s_{\rm BV}$ that characterizes the temporal evolution of the B-V color. The Phillips relation captures the behavior that brighter SNe Ia evolve more slowly (i.e. they have smaller $\Delta m_{15} (\rm B)$ and higher SALT $x_1$ with respect to the mean values of these parameters), while dimmer events decline more rapidly.

Although the observed data of Type Ia SNe show remarkable uniformity at first glance, significant diversities have been identified over the past decades. In fact, various characteristics can be used to define subclasses and subtypes within the general SN Ia population. 

Studying the distribution of the Si II $\lambda6355$ line velocity near maximum brightness, \cite{wang09} noted a bimodal tendency, and introduced two subtypes based on this. Normal velocity (NV) events have $v_{\rm Si}(t_{\rm max}) < 11,800$ km~s$^{-1}$, while high velocity (HV) events show $v_{\rm Si}(t_{\rm max}) > 11,800$ 
km~s$^{-1}$. In this paper, we adopt a threshold of $12000$ km~s$^{-1}$. HV SNe Ia also tend to have redder colors \citep{wang09, mandel14} at maximum, and often show weaker or absent C II $\lambda6580$ features at early phases compared to NVs \citep{li21}. Among other things, viewing angle effects, differences in progenitor systems, local environment, or explosion mechanisms have all been proposed to explain the separation of the two groups \citep[e.g.][]{pan24}.

By looking at the post-maximum gradient of the Si II $\lambda6355$ velocity ($\dot{v}$), \citet{benetti05} also defined three distinct groups of SNe Ia. Low velocity gradient (LVG, $\dot{v} < 70 \text{ km s}^{-1}\text{d}^{-1}$) SNe are brighter, slower events that include 91T-likes, while high velocity gradient (HVG, $\dot{v} > 70 \text{ km s}^{-1}\text{d}^{-1}$) objects represent the normal SNe Ia. The additional Faint group consists of dim, 91bg-like events exhibiting lower, rapidly evolving Si II velocities.

\citet{branch06} used the pseudo-equivalent widths (pEWs) of the Si II $\lambda6355$ and $\lambda5972$ features to differentiate between four groups. Core-normal (CN) SNe Ia have typical spectral features, and they are associated with photometrically normal events; shallow-silicon (SS) SNe Ia show weaker Si lines and often include 91T-like slow-decliners; broad-line SNe (BL) have broader spectral features; and cool (CL) events generally represent faint, 91bg-like fast-decliners.

The main focus of this paper is on the high-velocity features (HVFs; not to be confused with the HV category of \cite{wang09}) of SNe Ia. HVFs were first described in the early-phase spectra of SN~1994D by \cite{hatano99}. They were able to reproduce the spectrum only by adding additional  lines of Fe II and Ca II to their model, with velocities much higher than that of the other (photospheric) lines. HVFs are detected most often in the near-infrared triplet (NIR3, $\lambda 8579$ \AA) as well as the H\&K lines ($\lambda 3945$ \AA) of Ca II, but they also frequently appear in the Si II $\lambda6355$ feature at early phases \citep{mazzali05b, harvey25}. Occasionally, HVFs were identified in S~II, Fe~II \citep{marion13} and O~I features as well \citep{zhao16, zhao24}. They appear at higher velocities than their photospheric counterparts (the latter called photospheric-velocity features, PVFs, hereafter) by $\sim 6000 \text{ km s}^{-1}$, usually fade soon after maximum, and show significant polarization compared to the continuum of the SN Ia spectrum \citep{kasen03, maund13}. Their velocity and polarization indicates that these features may be formed in an asymmetric region further out from the photosphere, perhaps in multiple blobs of material or in a torus \citep{thomas04, branch04, tanaka06}, or shell. The presence of this asymmetric or clumpy line-forming region could be either due to a local abundance enhancement, a density enhancement, or ionization effects  \citep[e.g.][]{mazzali05a}, among other things.

Previous studies on large samples revealed correlations with other observable properties of SNe Ia: the strength and average velocity of these HVFs tend to decrease with the LC decline rate, and they are very rare in 91bg-like fast decliners \citep{childress14, maguire14}. SNe Ia with lower Si II $\lambda6355$ velocity at maximum also seem to show stronger HVFs, while in the Wang HV SNe Ia they fade quickly, and are generally weaker or even absent \citep{silverman15}. Their host galaxy environment also seems to have an effect on the nature of HVFs: SNe in late-type, young, high specific star formation rate (sSFR) hosts tend to show stronger HVFs \citep{pan15,silverman15,meng19}.

Despite intensive observational study, the physical origin of the HVF remains unclear. The question of the origin of HVFs is closely tied to our fundamental (and so far unanswered) questions about the progenitors and explosion mechanism(s) of Type Ia SNe as a whole.

This paper is structured in the following way. In Section \ref{sec:obs} we detail how the data we used was acquired and provide a list of the studied objects. In Section \ref{sec:method}, we describe the methods and techniques that were used to derive line velocities and other spectral parameters from the spectra. Section \ref{sec:results} describes the results we obtained from studying the evolution and  connection of spectral parameters with other characteristics of the SNe. Finally, in Section \ref{sec:discussion}, we discuss the possible implications of our results, and compare them with previous studies. Section \ref{sec:conclusions} provides the conclusions and summary of our work.

\section{Observations}
\label{sec:obs}

The SN group at UT Austin has been collecting early optical spectra on SNe Ia with the Hobby-Eberly Telescope \citep[HET;][]{ramsey98} at McDonald Observatory for nearly two decades. The purpose of this program was to capture the appearance and evolution of HVFs in the earliest spectra of SNe Ia, and to reveal the diversity of HVFs in different subtypes. Spectra were taken with the Marcario Low Resolution Spectrograph \citep[LRS;][]{hill98} between 2008 and 2013. After the HET wide-field upgrade \citep{hill21}, the observations were resumed with the Low Resolution Spectrograph 2 \citep[LRS2;][]{chonis16} in 2016 and have been continuing since then.

The LRS spectra covered the 4400--9200 \AA\ wavelength region with a resolving power  of $R \sim 400$ at 6000 \AA . These spectra did not extend to the blue to contain the Ca II H\&K feature around 3945 \AA , but the Ca II NIR3 feature around 8579 \AA\ was fully covered. The data were reduced by standard IRAF scripts, including bias subtraction, flatfield division and telluric removal. Wavelength- and flux calibrations were done using CdHg and ArNe spectral lamp exposures and observations of spectrophotometric standard stars, respectively. 

LRS2 is a fiber-fed, two-beam spectrograph containing two Integral Field Units (IFU), LRS2-B and LRS2-R, both having 660 densely packed fibers covering a 12"$\times$6" field-of-view. The LRS2-B beam is split into two arms: the UV-arm covers the 3700-4700 \AA\ region with a resolving power of $R \sim 1910$, while the Orange-arm extends from 4600 to 7000 \AA\ having $R \sim 1140$. LRS2-R has a Red-arm from 6500 to 8470 \AA\ with $R \sim 1760$, and a Far-red arm between 8230 and 10500 \AA\ with $R \sim 1920$. The spectra are reduced and calibrated automatically by the {\tt Panacea}\footnote{\url{https://github.com/grzeimann/Panacea}} pipeline running at the Texas Advanced Supercomputer Center (TACC\footnote{\url{https://tacc.utexas.edu/portal/login?next=/portal/}}). {\tt Panacea} produces 3D data cubes having $33 \times 20 \times 2064$ pixel resolution for each arm. The final spectrum extends from 3700 to 10500 \AA, and it covers both the Ca II H\&K and the NIR3 features. 

Between 2009 and 2024, spectra on 269 different SNe have been collected by HET. Among them 182 were classified as SN Ia, but only about 1/3 of them were observed early enough to detect HVFs in their spectra. After visual inspection, 56 SNe Ia were selected for further analysis. Many of them were observed multiple times, resulting in 125 HET spectra in total. The full set of HET spectra, including other types of SNe and transients, will be publicly released in a separate paper.

\subsection{Supplementary data}

To achieve better phase coverage, we also collected publicly available spectra from the Transient Name Server\footnote{\url{https://www.wis-tns.org/}}. Information about all the spectra used here, their date of observation, wavelength coverage and source can be found in the data repository of this paper\footnote{\dataset[https://doi.org/10.5281/zenodo.22079031]{https://doi.org/10.5281/zenodo.22079031}}, as well as in the Appendix.

For estimating the phase of our spectra, knowledge of the time of maximum brightness ($t_{\rm max}$) was needed. When it was available, we used previously published values for $t_{\rm max}$, as seen in the relevant column of Table \ref{tab:sample}. When the necessary information was not available in the literature, we used SNID \citep{blondin07} to estimate the phase of the spectrum and approximated the time of maximum brightness from that. In some other cases, we also looked at the ZTF \citep{bellm19, alerce} or ATLAS \citep{smith20} light-curves to help estimate $t_{\rm max}$. 

For 9 objects (SN~2021gtp, SN~2021hiz, SN~2022erw, SN~2022zut, SN~2023bee, SN~2023pbe, SN~2023tky, SN~2023vyl, SN~2024gy), photometric data were available from the RC80 telescope at the Piszkéstető Station of Konkoly Observatory, Hungary. By fitting their light-curves with the SALT3 model \citep{kenworthy21}, we were able to obtain a reliable time of maximum for these objects, along with their SALT $x_1$ parameters (Table \ref{tab:phys_data}.). For 5 objects, these parameters were already estimated by \cite{bora24}, but the last 4 were freshly fit for this paper.

\subsection{Sample}
The final sample consists of 56 SNe Ia  with 181 optical and near-infrared spectra that show HVFs in at least the Ca II NIR triplet (referred to as Ca II NIR3 in this paper). Our sample does not contain SNe without any HVF, so it is not representative of the entire SNe Ia population. The basic information of our objects is listed in Table~\ref{tab:sample}. 

\begin{table*}[ht]
    \tiny
    \centering
    \setlength{\tabcolsep}{2pt}
    \begin{tabular}{lllccc|lllccc}
    \hline
    \hline
    Name     &   Host galaxy   &   z   &   $t_{\rm max}$   &   Type  & Refs.  &  Name     &   Host galaxy   &   z   &   $t_{\rm max}$   &   Type  & Refs.\\
             &       &     &   (MJD)   &    &    &      &       &     &   (MJD)   &    & \\
    \hline
    SN~2009ig       &  NGC 1015 &  0.0088 & 55080  &  Ia-n  & d  & SN~2013dy      &  NGC 7250 &  0.0039 & 56501  &  Ia-n/pec\\
PTF~10bjs      &  MCG +09-21-083 &  0.0300 & 55261  &  Ia-n   & d    &   SN~2017cbv       &  NGC 5643 &  0.0040 & 57841  &  Ia-n/pec & f \\
PTF~10icb       &  CGCG 293-042 &  0.0088 & 55360  &  Ia-n  & d  &   SN~2017hjy       &  SDSS J023602.15+432817.6 &  0.0070 & 58056  &  Ia-n  & i \\
PTF~10qjl       &  WISEA J163958.56+120624.0 &  0.0579 & 55419 &  Ia-n  & c, d   &   SN~2017ln      &  LEDA 1735060&  0.0254 & 57777  &  - & \\
PTF~10ygu        &  NGC 2929 &  0.0250 & 55500  &  Ia-n  & d     &   SN~2018gv     &  NGC 2525 &  0.0053 & 58149   &  Ia-n  & h, j\\
SN~2010ai       &  LEDA 126792 &  0.0183 & 55277  &  Ia-n  & d   &   SN~2018ilu      &  SDSS J233320.80+044839.0 &  0.0179 & 58454  &   -  &\\
SN~2010dm      &  LEDA 5068074 &  0.0315 & 55349  &  99aa  & d  &   SN~2018oh      &  UGC 4780 &  0.0110 & 58163  &  Ia-n & i \\
SN~2010it       &  NGC 1417 &  0.0138 & 55490  &   Ia-n  & d     &   SN~2020aabz      &  AGC 1256778 &  0.0263 & 59185  &  -  & \\
SN~2010iw       &  UGC 4570 &  0.0215 & 55492  &    99aa  & d, e    &   SN~2021J      &  NGC 4414 &  0.0024 & 59235 &   Ia-n  & i, l\\
SN~2010ke      &  WISEA J005724.60-005748.5 &  0.0420 & 55532  & - &    &   SN~2021gtp      &  UGC 4611 &  0.0199 & 59309  &  Ia-n & k  \\
SN~2010kg       &  NGC 1633 &  0.0166 & 55542  &    Ia-n  & d    &   SN~2021hiz      &  IC 3322A &  0.0033 & 59320  &  Ia-n & k, l\\
SN~2011ao       &  IC 2973 &  0.0107 & 55633  &    Ia-n  & d     &   SN~2022erw      &  IC 0651 &  0.0150 & 59657  &  -  & \\
SN~2011by      &  NGC 3972 &  0.0028 & 55690  &    Ia-n  & d, e      &   SN~2022hkc      &  UGC 3944 &  0.0130 & 59687  &  -  & \\
SN~2011dm       &  UGC 11861 &  0.0049 & 55737  &    Ia-n  & d   &   SN~2022yhl     &  LEDA 1126510 &  0.0856 & 59879  & - &  \\
SN~2011fe       &  M101 &  0.0008 & 55815 &    Ia-n  & d, i      &   SN~2022zut       &  NGC 3810 &  0.0033 & 59905  &  Ia-n & k \\
SN~2011fg       &  WISEA J232320.69+164742.5 &  0.0450 & 55790  &  Ia-pec &   &   SN~2023alb     &  WISEA J110741.22+565116.4 &  0.1400 & 59974  &  - & \\
SN~2011hb      &  NGC 7674 &  0.0290 & 55872  &    Ia-n  & d     &   SN~2023bee      &  NGC 2708 &  0.0067 & 59994   &  Ia-n & k  \\
SN~2012bh       &  UGC 7228 &  0.0252 & 56016   &    Ia-n  & d, g    &   SN~2023eoc      &  WISEA J161130.45+522745.5 &  0.0360 & 60054  &  -&  \\
SN~2012bm       &  UGC 8189 &  0.0248 & 56021   &   - & g   &   SN~2023ktw       &  WISEA J172730.78+341724.3 &  0.0700 & 60123  &  - & \\
SN~2012cg      &  NGC 4424 &  0.0015 & 56018   &  91T/99aa  & d, g      &   SN~2023pbe     &  Z 427-27 &  0.0202 & 60181  &  - &  \\
SN~2012da      &  GMP 1396 &  0.0240 & 56119  &   Ia-n  & d      &   SN~2023tky     &  LEDA 138116 &  0.0466 & 60221  &  - & \\
SN~2012fr     &  NGC 1365 &  0.0055 & 56243  &  Ia-n/91T & a     &   SN~2023vyl     &  NGC 7625 &  0.0054 & 60260  &  - & \\
SN~2012G     &  IC0803NED01 &  0.0259 & 55951  &   - &      &   SN~2023xtf      &  LEDA 2129246 &  0.0355 & 60288  &  - & \\
SN~2012hj     &  MCG +08-20-89 &  0.0246 & 56276  &  - &        &   SN~2023xuz    &  LAMOST J214816.85-020404.5 &  0.0500 & 60274  &  - & \\
SN~2012ht     &  NGC 3447 &  0.0036 & 56295  &   Ia-04eo  & b   &   SN~2023zvn      &  LEDA 31891 &  0.0480 & 60305  &  - & \\
SN~2013ag     &  SDSS J125135.41+263744.0 &  0.0213 & 56357  & - &      &   SN~2024gy      &  NGC 4216 &  0.0004 & 60329  &  Ia-n  & m  \\
SN~2013be      &  IC 3573 &  0.0659 & 56398  &  Ia-n  & g    &   SN~2024iei      &  LEDA 2196050 &  0.0549 & 60445  &  - & \\
SN~2013di      &  NGC 7321 &  0.0238 & 56469  &   Ia-n  & d      &   SN~2024xyn      &  2MFGC 02100 &  0.0215 & 60608  &  - &  \\

    \hline
    \end{tabular}
    \caption{Basic data of the supernovae making up our sample: their names, most likely host galaxies, redshifts, time of maximum brightness and subtype. When the subtype is unknown, the column is left empty, while ``Ia-n'' marks the normal SNe Ia. The last column lists the publications containing the used information.\\ (a):\cite{childress13}, (b):\cite{yamanaka14}, (c): \cite{maguire14}, (d): \cite{silverman15}, (e): \cite{friedman15}, (f): \cite{hosseinzadeh17}, (g): \cite{weyant18}, (h): \cite{stahl19}, (i): \cite{ktr20}, (j): \cite{yang20}, (k): \cite{bora24}, (l): \cite{ma25}, (m): \cite{li26}}
    \label{tab:sample}
\end{table*}

Out of these 56 events, 40 SNe show signs of an HVF in the Si II $\lambda$6355 \AA~line as well, in at least one spectrum. 21 objects have a C II $\lambda$6580 \AA~feature present in their spectra at early times, and among these, 2 of them seem to have an HVF in this carbon feature as well. We identify 12 SNe showing a detectable Na I D absorption in our sample.

\begin{table*}[ht]
    \centering
    \footnotesize
    \begin{tabular}{lcccc|lcccc}
    \hline
    \hline
    Name    &  Wang subtype&  Si II HVF &  C II &   Na I D & Name    &  Wang subtype &  Si II HVF&  C II &   Na I D\\
    \hline
    SN~2009ig     & HV & yes  & yes & yes & SN~2013dy     & NV & yes  & yes & - \\
    PTF10bjs     & HV & - & - & - & SN~2017cbv     & NV & yes  & HVF & yes \\
    PTF10icb     & NV  & yes & yes & yes & SN~2017hjy     & NV & yes  & - & - \\
    PTF10qjl     & NV & -  & - & - & SN~2017ln     & NV & -  & - & - \\
    PTF10ygu     & NV & yes  & - & - & SN~2018gv     & NV & yes  & yes & - \\
    SN~2010ai     & NV & yes  & yes & - & SN~2018ilu     & NV & yes  & yes & - \\
    SN~2010dm     & NV & yes & - & - & SN~2018oh     & NV & yes  & yes & - \\
    SN~2010it     & HV & yes & - & - & SN~2020aabz     & NV & -  & - & - \\
    SN~2010iw     & NV & yes  & - & - & SN~2021J         & NV & -  & - & yes \\
    SN~2010ke     & NV & yes  & - & - & SN~2021gtp     & NV & -  & - & - \\
    SN~2010kg     & NV & yes  & - & - & SN~2021hiz     & NV & yes  & yes & - \\
    SN~2011ao     & NV & yes & - & -  & SN~2022erw     & HV & -  & - & yes \\
    SN~2011by     & NV & yes  & yes & yes & SN~2022hkc     & NV & -  & - & - \\
    SN~2011dm     & HV & yes  & yes & yes & SN~2022yhl     & HV & yes  & - & - \\
    SN~2011fe     & NV & -  & yes & - & SN~2022zut     & HV & -  & - & - \\
    SN~2011fg     & NV & -  & - & - & SN~2023alb     & NV & yes  & - & - \\
    SN~2011hb     & NV & yes  & - & yes & SN~2023bee     & NV & yes  & HVF & - \\
    SN~2012bh     & NV & -  & - & - & SN~2023eoc     & NV & yes  & - & - \\
    SN~2012bm     & NV & yes  & - & yes & SN~2023ktw     & NV & yes  & - & - \\
    SN~2012cg     & NV & yes  & yes & yes & SN~2023pbe     & NV & yes  & - & - \\
    SN~2012da     & NV & yes  & yes & - & SN~2023tky     & NV & yes  & - & - \\
    SN~2012fr     & NV & yes  & yes & - & SN~2023vyl     & NV & -  & - & - \\
    SN~2012G     & NV & yes  & - & - & SN~2023xtf     & NV & yes  & - & - \\
    SN~2012hj     & NV & yes  & yes & - & SN~2023xuz     & NV & yes  & - & - \\
    SN~2012ht     & NV & -  & yes & -  & SN~2023zvn     & NV & -  & - & - \\
    SN~2013ag     & NV & -  & - & - & SN~2024gy     & NV & yes  & yes & yes \\
    SN~2013be     & NV & yes  & - & - & SN~2024iei     & NV & yes  & - & - \\
    SN~2013di     & NV & yes  & yes & - & SN~2024xyn     & NV & yes  & yes & yes \\
 
    \hline
    \end{tabular}
    \caption{Spectral characteristics of the supernovae making up our sample: their \cite{wang09} subtype, and whether or not they show evidence for Si II HVF, C II, or Na I D absorption. The two instances where an HVF was detected in C II are marked with ``HVF''.}
    \label{tab:sample2}
\end{table*}

\section{Methodology}
\label{sec:method}

After correcting the spectra for redshift (collected from \cite{ned} for each host galaxy), we used two methods to determine the velocities of the Si II $\lambda$6355 \AA, the Ca II NIR triplet, and the Ca II H\&K spectral features. First, we fitted multiple Gaussians to the line regions, and used the minima of the fitted functions to calculate the velocities. Second, we utilized the SYNOW spectral modeling code \citep{fisher99} and fitted the observed spectra with models by eye to find the velocities, line optical depths, and the width of the line forming region for the photospheric and high-velocity lines.
We compare the velocities acquired from the Gaussian fitting to those determined from spectral modeling in Section \ref{sec:gauss_vs_synow}.

\subsection{Gaussian fits}
\label{sec:gauss}

We used Gaussian fits to determine the HVF and PVF velocities of the Si II $\lambda$6355 \AA, the Ca II NIR triplet and the Ca II H\&K features.

For the Si II $\lambda$6355 \AA, we used a function with two Gaussians (one for the HVF, one for the PVF), where the fitted parameters were
\begin{itemize}
    \item the central wavelength of the high-velocity feature,
    \item the separation (in \AA) of the photospheric line center from the HVF,
    \item the amplitude and width of the HVF and PVF,
    \item a shift along the y-axis.
\end{itemize}
We gave initial guesses and bounds for each of these parameters by hand, making sure that the separation between HVF and PVF always remained $>0$ \AA. In the cases where the HVF was no longer visible, we only used a single Gaussian to determine the velocity of the photospheric component.

For the case of the Ca II H\&K and NIR3 lines, the fitting procedure was largely the same, with only a few complications. Since the H\&K and NIR3 lines are themselves made up of 2 and 3 different lines, respectively, the two main Gaussians (representing the HVF and PVF) were taken to be a combination of multiple Gaussians. The central wavelengths of these ''sub-Gaussians`` were fixed relative to each other based on their rest wavelengths, and it was the strength-weighted mean wavelength of the line complex (3945 \AA~for the H\&K, and 8579 \AA~for the NIR3) which was later varied to find the best fit. The amplitudes of the individual lines were also fixed relative to each other based on their strengths.

\begin{figure*}[ht]
    \centering
    \includegraphics[width=0.95\linewidth]{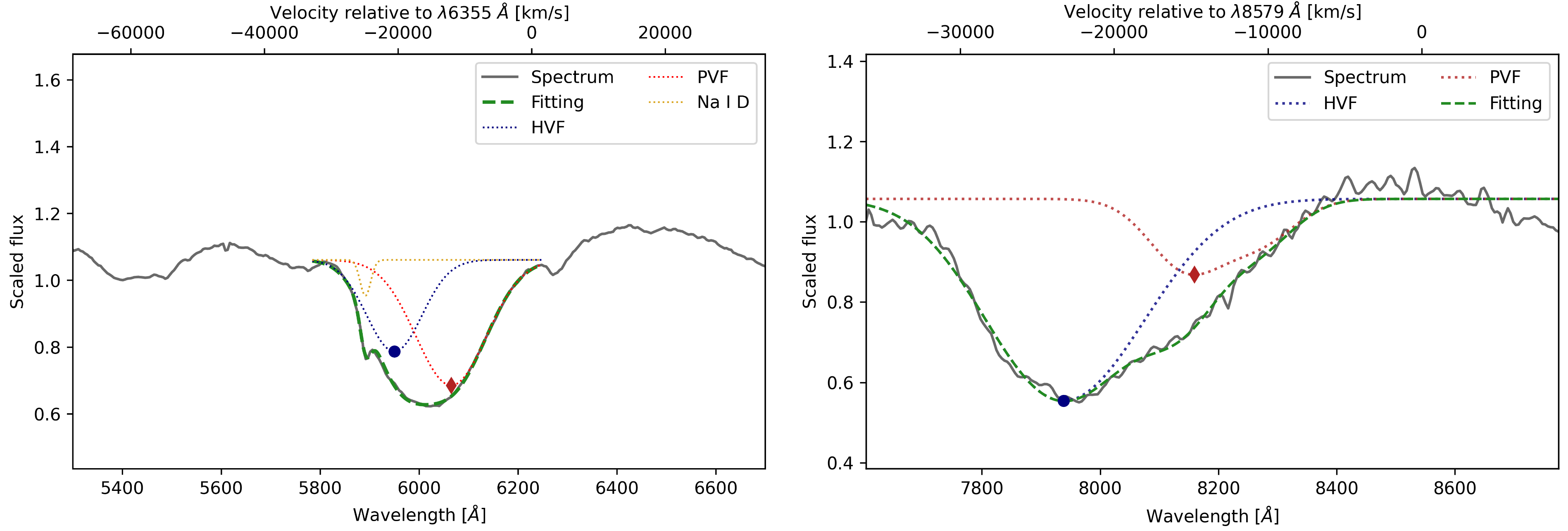}
    \caption{Spectra of SN~2012cg 14 days before B band maximum (solid gray line). Left: Gaussian fitting of the Si II $\lambda$6355 \AA~line when an HVF (blue dotted line) and the Na I D line (yellow dotted line) are also present. Right: Fitting of the Ca II NIR triplet in the same spectrum with an HVF (blue dotted line) and PVF (red dotted line). The dashed green lines on both plots show the sum of the individual Gaussian profiles.}
    \label{fig:gaussfit_examples}
\end{figure*}

When present in the spectrum, we also added a single Gaussian to the determine the velocity of the Na I D line. Because of its proximity to the Si II feature, sometimes the Na I D line blends with the Si II line. In such cases Na I D was included in the fitted function to achieve a better fit of the the Si II profile. One such example, along with a Ca II NIR triplet fitting is shown in Figure \ref{fig:gaussfit_examples}.

\begin{figure}[ht]
    \includegraphics[width=0.95\linewidth]{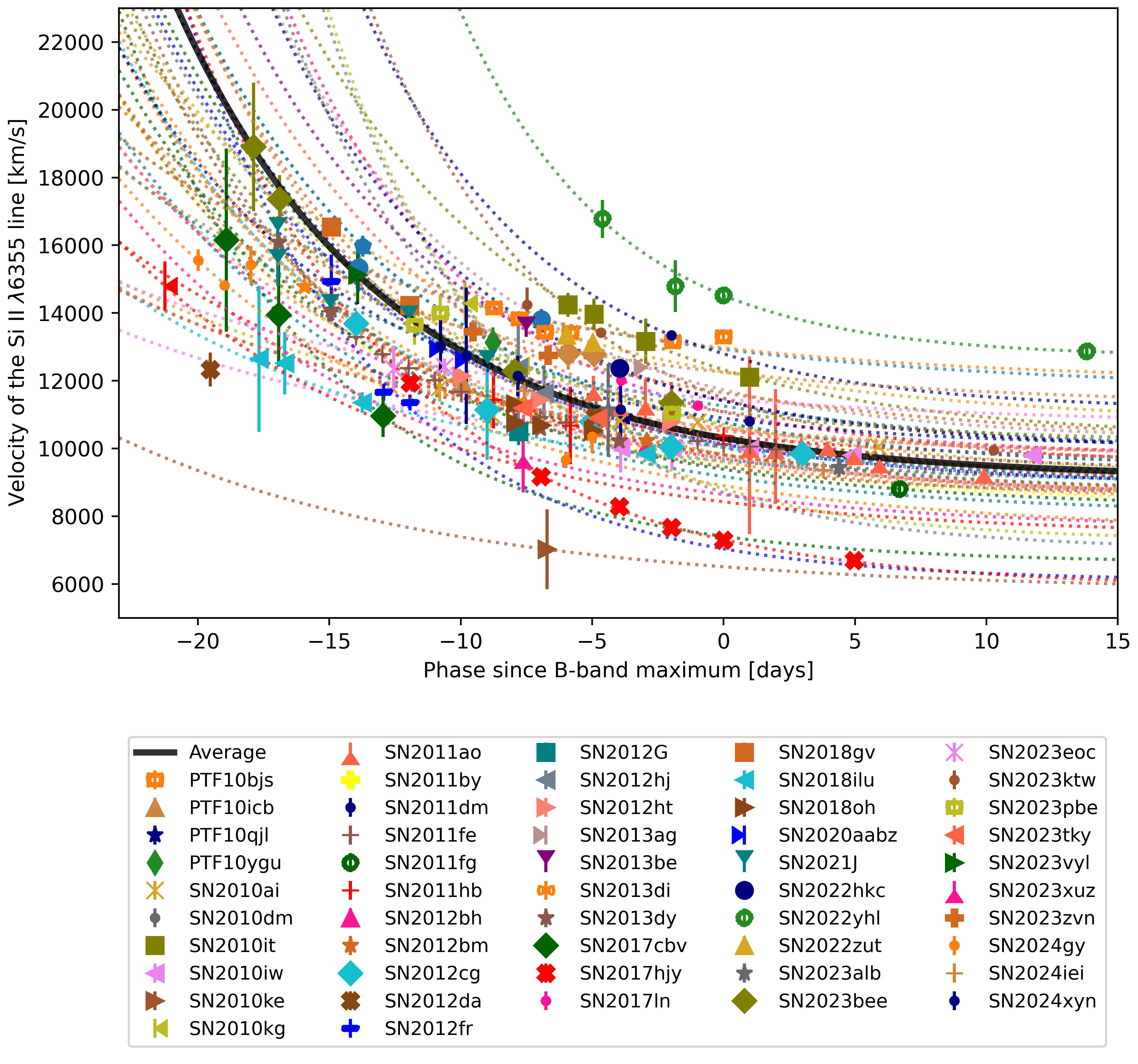}
    \caption{Si II $\lambda$6355 \AA~velocities determined from Gaussian fitting. The dotted lines show the fitted exponential functions that were used to estimate the photospheric velocity at the time of maximum light. All SNe are shown with different colors and markers. The solid black line shows the average of all the fitted functions, highlighting the general trend of the Si II velocities for the whole sample.}
    \label{fig:si_vels}
\end{figure}

The velocity of the Si II $\lambda$6355 \AA~line around the time of maximum ($v_{\rm Si}^{\rm tmax}$) is often used as a proxy for photospheric velocity, and it is an important physical parameter for Type Ia SNe. Because our paper focuses on high velocity features, which appear at early times, we generally lack the spectra around phase 0 that is necessary to directly measure $v_{\rm Si}^{\rm tmax}$. To circumvent this, we fitted the velocity curves defined by the early Si II PVF velocities with an exponential function and took the value of this function at the time of maximum as an estimate for $v_{\rm Si}^{\rm tmax}$. The measured Si II velocities and the functions fitted to them are shown in Figure \ref{fig:si_vels}, along with the average velocity function of the whole sample (solid black curve).

In 59 spectra of 21 SNe, a C II $\lambda$6580 \AA~absorption feature can be clearly seen near the P Cygni peak redward of the Si II absorption minimum. The presence of C II absorption could be a sign of unburned material in outer regions of the ejecta \citep{maguire14}. However, the unambigous detection of this carbon feature is difficult due to its proximity to the Si II $\lambda$6355 feature. Often the C II absorption is only visible as an apparent flatness of the P Cygni emission redward of the Si II absorption, and in extreme cases it may even blend together with the Si II absorption component \citep{parrent11}. We only calculated C II velocities for those cases when a clear C II absorption component was visible and clearly separated from Si II. Consequently, we most likely underestimated the number of objects in our sample having carbon absorption features in their spectra. For those cases when C II was unambiguously present, we determined the velocity of the C II $\lambda$6580 feature from the Doppler shift of the absorption minimum. It was found that, when present, the carbon lines are the strongest in the earliest phases, and their velocities follow the Si II PVF velocities closely, until they eventually fade around maximum light. In the two cases when this C II line showed an HVF, it only appeared in the earliest spectra, had a velocity comparable to that of Si II HVFs and faded quickly over a few days. We were not able to fit these carbon HVFs with Gaussian functions, so their velocity is only determined by SYNOW modeling (see Section \ref{sec:synow}).

\begin{figure*}[ht]
    \centering
    \includegraphics[width=0.78\linewidth]{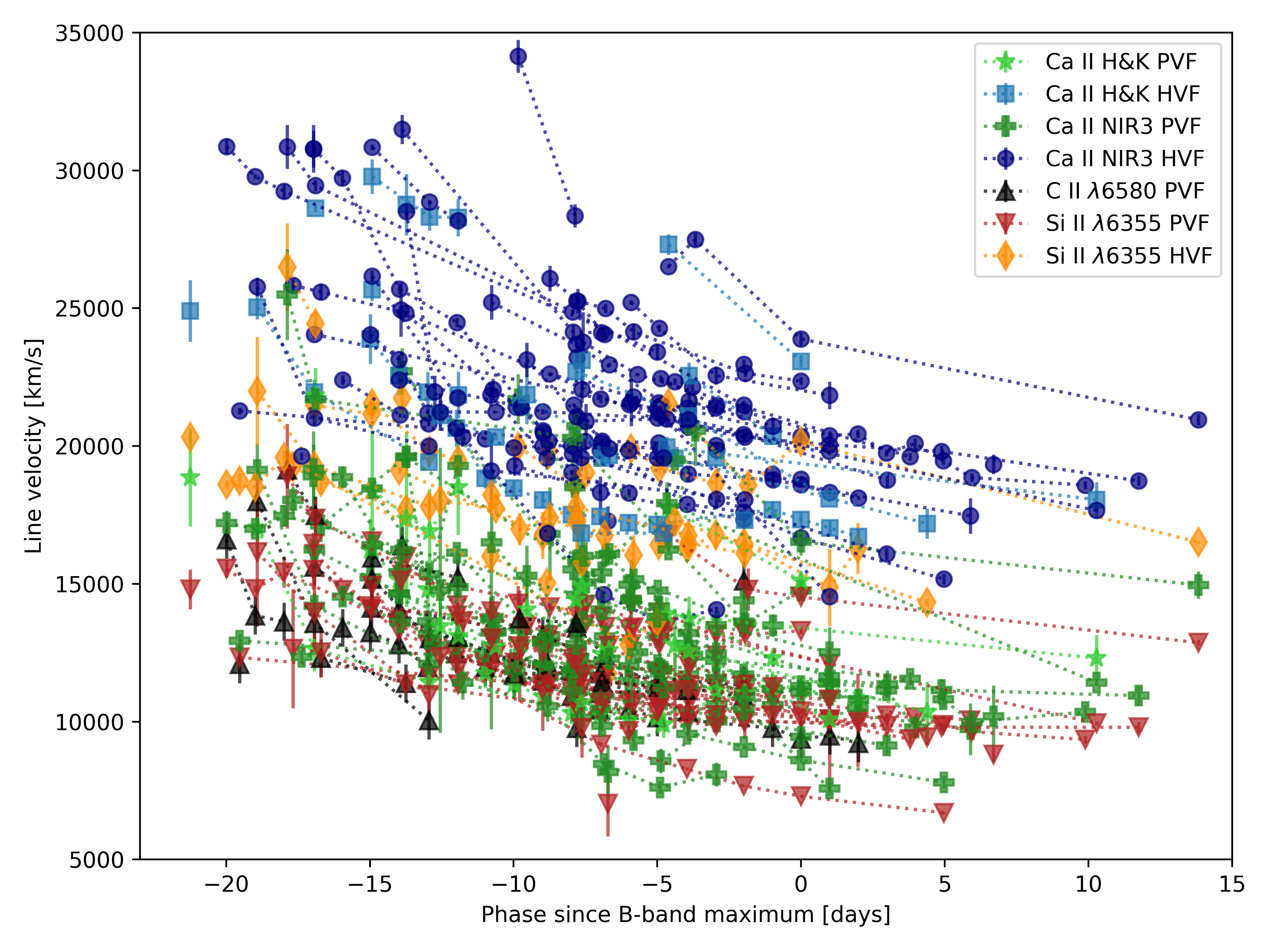}
    \caption{Results of the Gaussian fitting of spectral lines. Red triangles show the Si II photospheric velocities, and yellow diamonds their high-velocity components. Green stars and crosses denote the PVF and blue squares and dots the HVF velocities for the Ca II H\&K and NIR3 lines, respectively. Photospheric carbon is marked by black triangles. High-velocity carbon was identified in two cases, but only through SYNOW modeling, so they are not marked here.}
    \label{fig:v_all}
\end{figure*}

In Figure \ref{fig:v_all} we show all the velocities obtained by Gaussian fitting for the different lines, and their high velocity components, when present. 

Out of the 181 spectra used in this paper, 169 were in the wavelength range of the NIR triplet, while only 48 covered the H\&K doublet. Gaussian fitting of the Ca II H\&K feature can be problematic because of the presence of the Si II $\lambda$3585 \AA~line \citep{childress13}. Because of these problems, later we will focus only on the properties of the Ca II NIR triplet.

\begin{figure}[ht]
    \centering
    \includegraphics[width=0.95\linewidth]{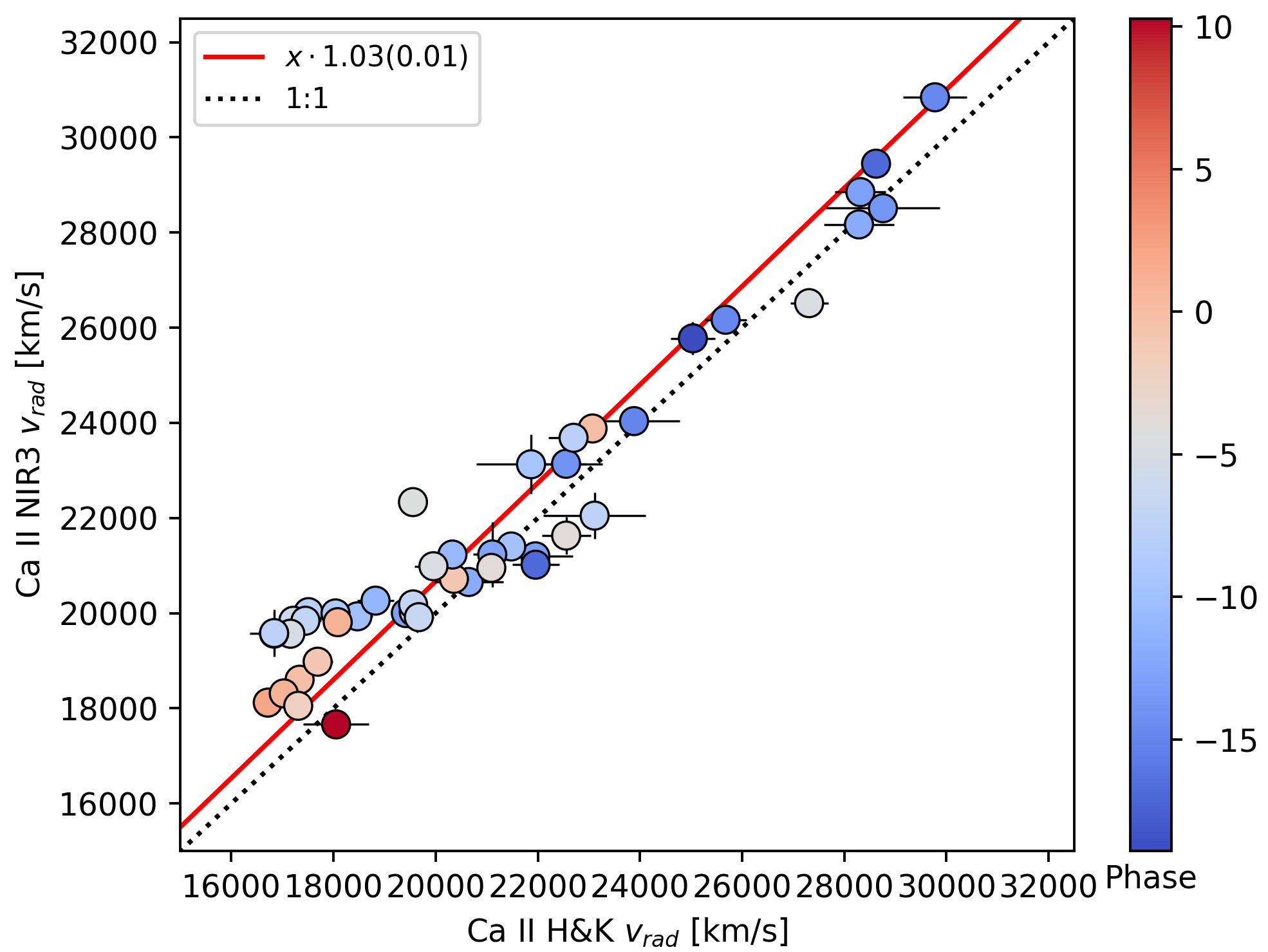}
    \caption{Comparison of the HVF velocities acquired by Gaussian fitting of the NIR triplet (vertical axis) and the H\&K lines (horizontal axis). Although the NIR3 appears to systematically have slightly higher velocities than the H\&K, the relation fitted to the data (shown with the red line) still follows the 1:1 relation shown with the black dotted line. The points are colored according to the phase (days from B band maximum).}
    \label{fig:hk_nir3}
\end{figure}

In the 45 cases when a spectrum did cover both the Ca II H\&K and NIR triplet, we compared their HVF velocities (see Figure \ref{fig:hk_nir3}). It is found that they are in good agreement with each other, as also shown earlier e.g. by \cite{zhao24}. This is a strong indication that these HVFs are not caused by misidentified other spectral features, or dominated by excitation states. Instead, they are indeed high-velocity lines belonging to Ca II. That being said, small differences between the H\&K and NIR3 velocities can be caused by the contamination of the  Si II $\lambda$3585 \AA~line close to the H\&K \citep{foley13}.

After finding reasonable Gaussian fits to the lines, we determined the pseudo-equivalent widths of both the HVF and PVF components, in order to calculate the strength of the HVFs. Following the methodology of \cite{childress14}, the strength of the HVF is defined as:
\begin{equation}
    R_{\rm HVF} = \frac{\text{pEW(HVF)}}{\text{pEW(PVF)}}.
    \label{eq:rpew}
\end{equation}

The parameter $R_{\rm HVF}$ has been widely used before in the literature to describe the strength of the HVFs, but it is important to point out a caveat of this method. Equation~\ref{eq:rpew} compares the pEW of the HVF to that of the PVF, so $R_{\rm HVF}$ actually describes the strength of the HVF \textit{compared to the PVF in the same spectrum}. Since the pEWs of PVFs also vary from SN to SN, this means that while $R_{\rm HVF}$ can be used to describe the evolution of the HVF in a single SN, for the comparison with the HVFs of other SNe is dubious. Despite this, we still use $R_{\rm HVF}$ for this purpose, but keep in mind the complications mentioned above.

When observing the evolution of the Ca II NIR3 HVF pEWs and $R_{\rm HVF}$, we see both properties clearly decreasing with time, leading to the fading of HVFs near maximum light.

Based on the temporal evolution of the calculated $R_{\rm HVF}$, we can find the phase $t_{\rm switch}$ at which the PVF starts to overpower the HVF (i.e. when $R_{\rm HVF}$ becomes $\leq1$). We were able to determine this parameter for the Ca II NIR triplet in 16 objects, and discuss its use in Section \ref{sec:csm}.

\subsection{Spectral modeling}
\label{sec:synow}
In addition to line fitting, we also constructed spectral models for each spectrum, and compared the velocities determined by these two methods.

\begin{figure}[ht]
    \centering
    \includegraphics[width=0.95\linewidth]{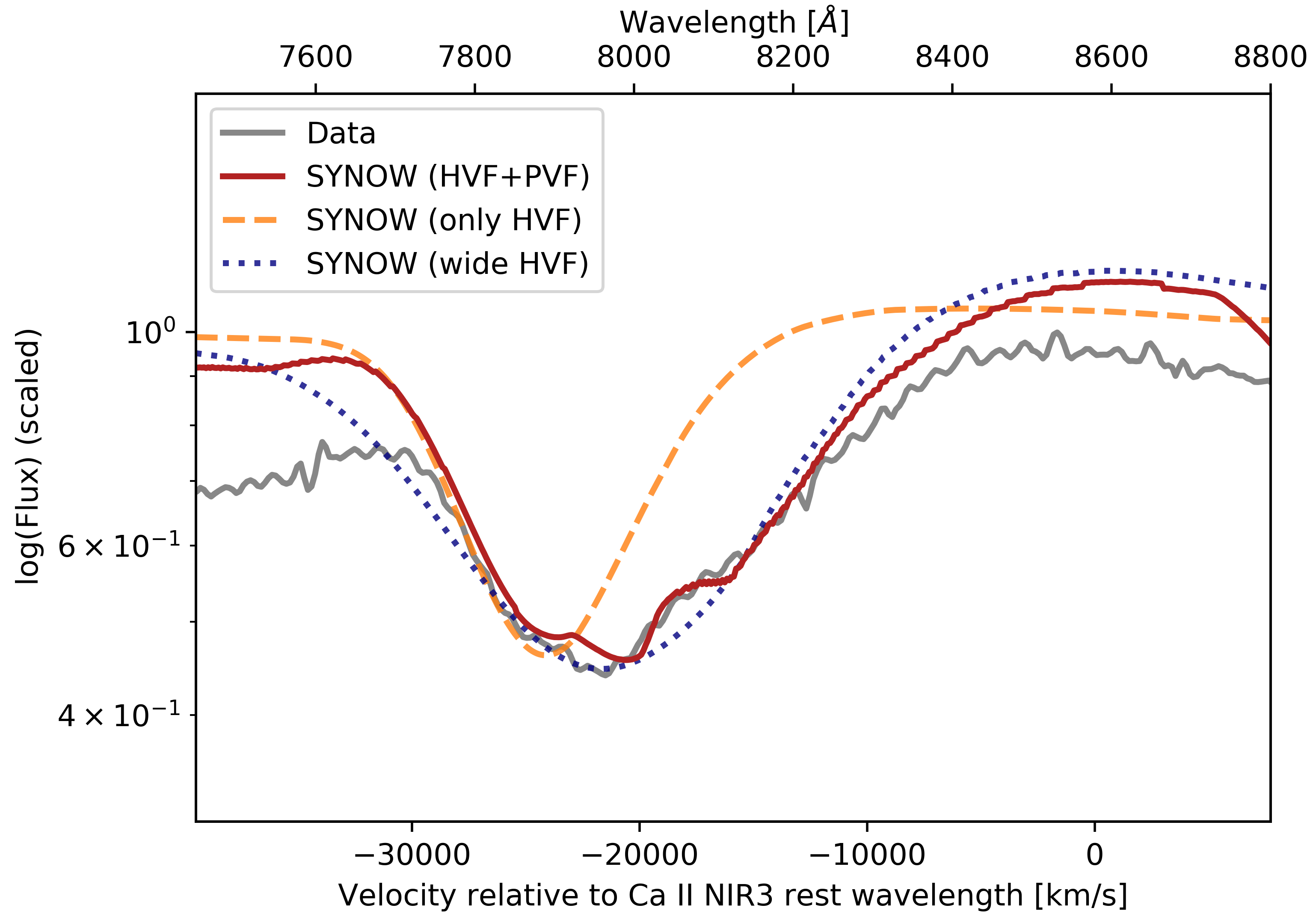}
    \caption{Ca II NIR3 line of SN2012cg 14 days before maximum light. The grey line shows the observed spectrum, and the red one the best-fit SYNOW model with both a PVF and HVF included. With the dashed orange line, we show a model where only a single Gaussian representing an HVF is present, and with the blue dotted line one where this Gaussian is particularly wide. Out of the 3 models, the one with both a PVF and HVF fits the observations best (red curve).}
    \label{fig:12cg_hvf}
\end{figure}

In Figure \ref{fig:12cg_hvf} we show that these features can only be fit reasonably when including both a photospheric and a detached, higher velocity component in the model. It is clear that a single spectral line (shown as an orange dashed line) alone cannot account for the observed features, no matter how wide it is (blue dotted line). Invoking both a photospheric and a high-velocity component together (plotted as the red solid line) describes the observed feature quite well. 

For the modeling, we used SYNOW, a parameterized supernova synthetic-spectrum code \citep{fisher99}, which allowed us to constrain the velocities of the HVFs relatively quickly. We included only Si II, Ca II, and sometimes C II in our models, as these were the features that showed (at least partly) resolved high-velocity components. After an initial guess, we varied the photospheric and HVF velocities ($v_{\rm phot}$, $v_{\rm HVF}$), line optical depths ($\tau_{\rm phot}$, $\tau_{\rm HVF}$) and line widths ($\sigma_{\rm phot}$, $\sigma_{\rm HVF}$) until a good fit to the observations could be achieved by eye. For reasons discussed previously, we concentrated on better modeling of the Ca II NIR3 feature rather than the H\&K lines.

Since our focus is the velocity of HVFs, we aligned the velocity at the photosphere to the PVF of the Ca II feature. Because Ca II lines are formed in a slightly higher velocity domain than the Si II features, this in some cases led to a small offset of the modeled Si II lines compared to the observations. In the cases of SN~2017cbv and SN~2023bee, we also included a high-velocity C II component in the earliest phases to model the strong absorption feature neighboring the Si II $\lambda$6355 \AA~line.

\begin{figure}[ht]
    \centering
    \includegraphics[width=0.95\linewidth]{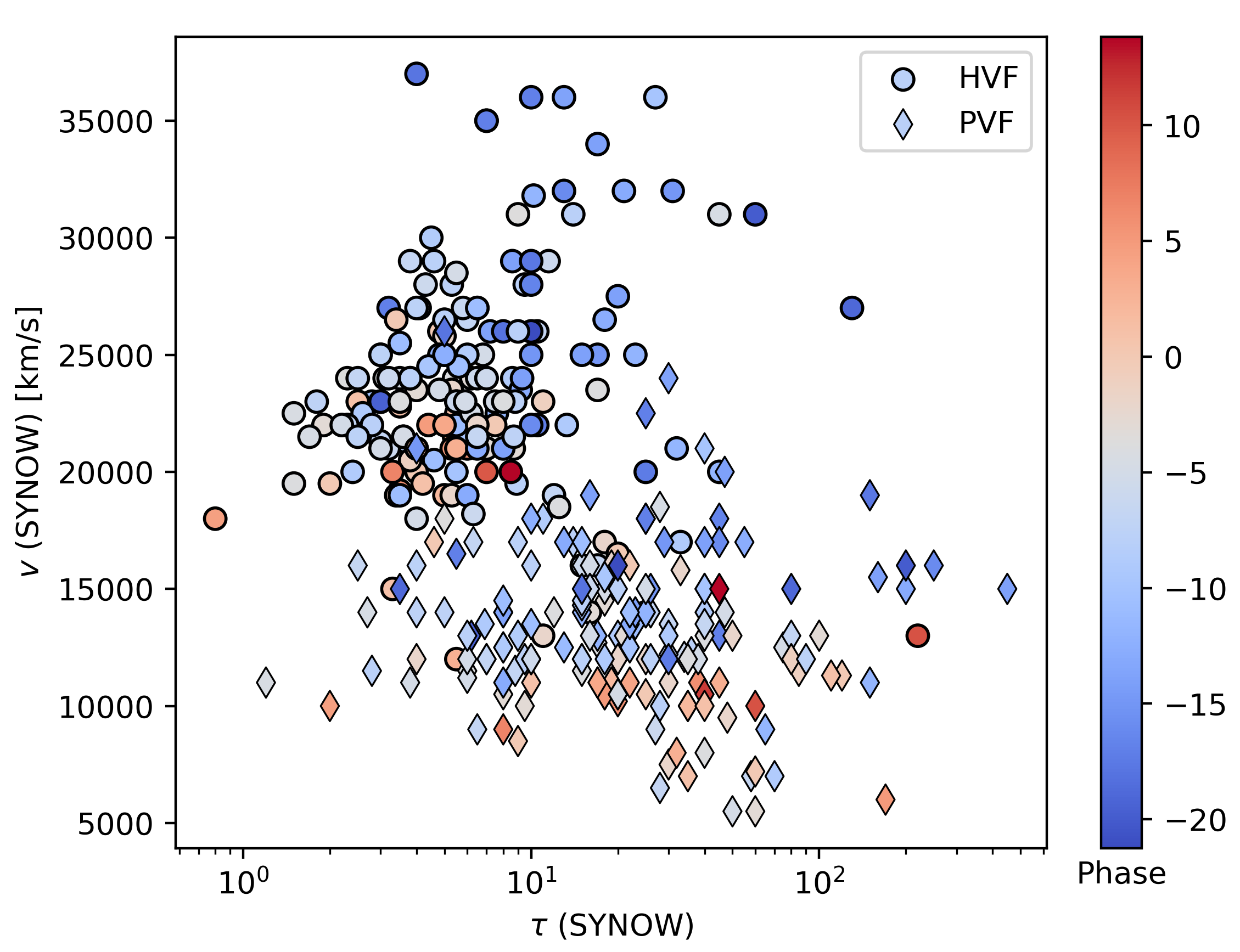}
    \caption{Line velocity of the HVF lines against their line optical depth for the Ca II components according to the SYNOW modeling. HVFs are shown with thick bordered circles, PVFs with thin diamonds. Coloring is according to phase, which generally increases from upper right to lower left. The two components are clearly separated and display different time-evolutions.}
    \label{fig:Ca_tau_v}
\end{figure}

Plotting the modeled Ca II HVF line optical depths against velocity for the whole sample shows a clear distinction between the photospheric and the high-velocity components (Figure \ref{fig:Ca_tau_v}). HVFs start strong and fast, and as the phase progresses, they reach lower velocities and line optical depths, until they eventually fade with\textit{out} blending into the PVFs. PVFs start at higher velocities and progress to lower speeds with time, but they do not appear to have a clear general evolution in their line optical depths. Based on this figure, we can see that the HVFs simply fade after maximum while maintaining a velocity difference with respect to the PVFs and do not disappear by slowing down and blending into the photospheric lines.

\begin{figure}[ht]
    \centering
    \includegraphics[width=0.98\linewidth]{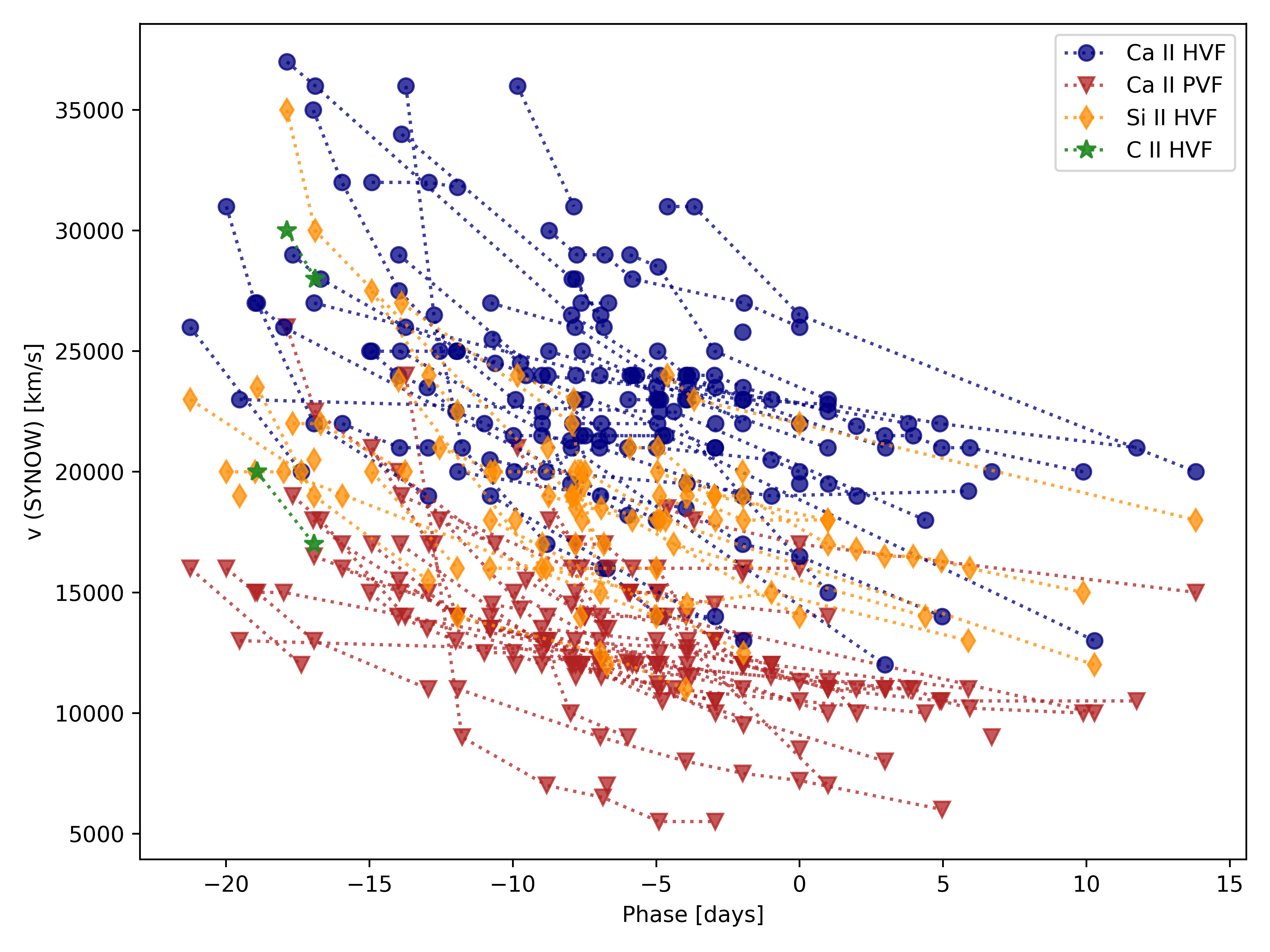}
    \caption{Velocities of the photosphere (aligned to the Ca II PVF) are plotted with red triangles. The HVFs of Si II, C II and Ca II are shown with yellow diamonds, green asterisks, and blue circles, respectively. The general velocity evolution of the features is similar to what we obtained from Gaussian fittings (Figure \ref{fig:v_all}), but differences between the 2 methods for each object can be discerned.}
    \label{fig:synow_v_all}
\end{figure}

The evolution of modeled velocities for the whole sample is shown in Figure \ref{fig:synow_v_all}. The photospheric and high-velocity areas are clearly distinct from each other for Ca II, but overlap a bit more in the case of Si II. This overlapping is most likely caused by two factors. As we said before, we adjusted the photospheric velocity to the Ca II PVF, which is typically higher velocity than the Si II PVF. Along with this, the HVF velocities tend to be lower for Si II than they are for Ca II (see Section \ref{sec:SiII_HVFs}), so the combined effects of lower Si II HVF velocities but higher photospheric velocities lead to the overlap of the Si II PVF and HVF velocity regions in Figure \ref{fig:synow_v_all}.

\subsection{Difference between Gaussian fits and spectral modeling}
\label{sec:gauss_vs_synow}
\begin{figure}[ht]
    \centering
    \includegraphics[width=0.93\linewidth]{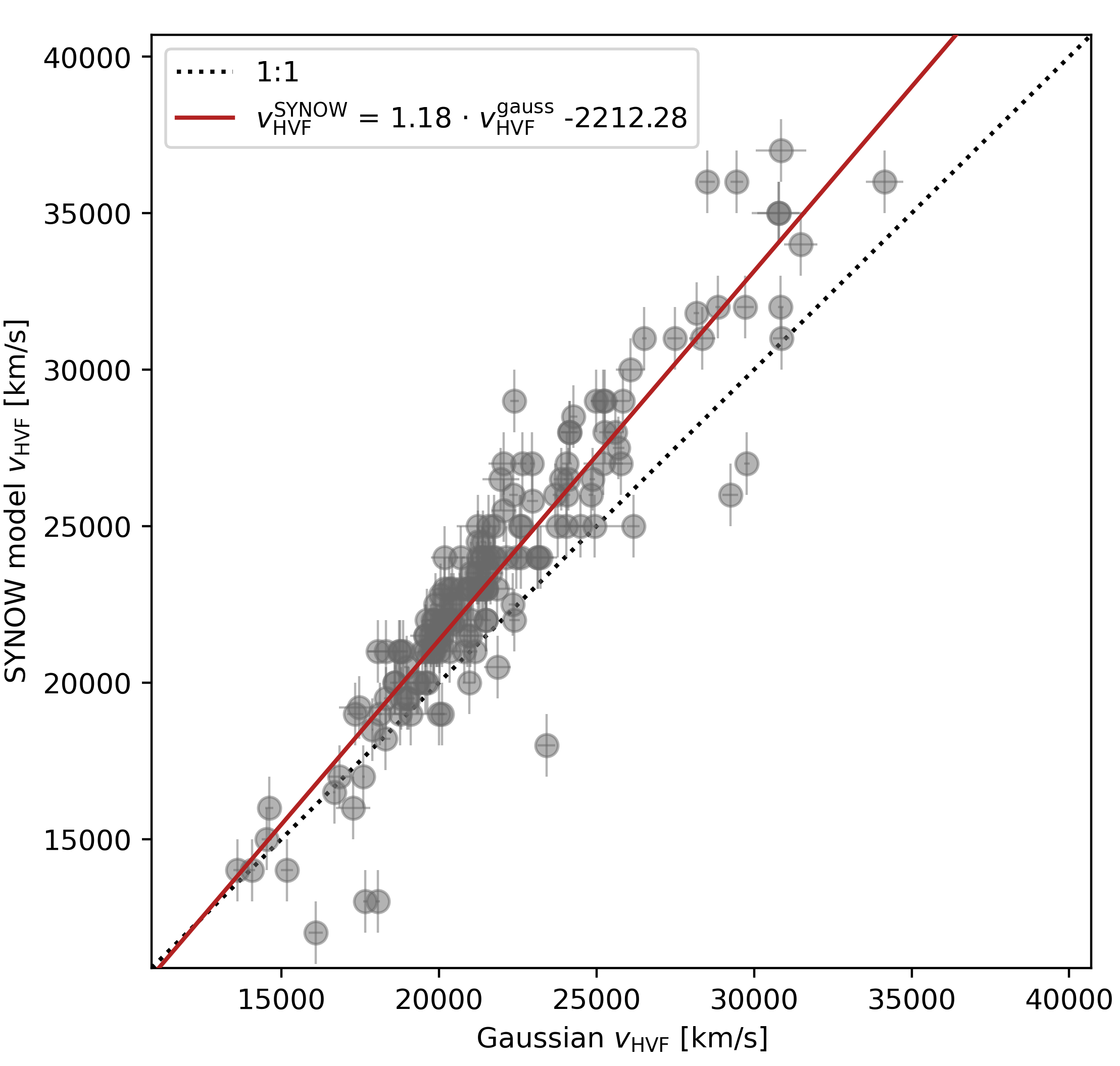}
    \caption{The SYNOW modeled velocities versus those determined by Gaussian fitting. The red line shows the empirical relation between the two. We assigned an uncertainty of $\pm1000$ km~s$^{-1}$ for all SYNOW velocities, as this was the uncertainty usually taken during our fit-by-eye process. The dotted line represents hypothetical equality between the two methods of velocity estimation.}
    \label{fig:gauss_vs_synow}
\end{figure}

When comparing the HVF velocities acquired from Gaussian fitting to those determined by SYNOW modeling, a difference of 500 to 2000 km~s$^{-1}$ can be discerned (referred to as $\Delta v_{\rm HVF}$ from now on), with the modeled velocities almost always being higher than the fitted ones (see Figure \ref{fig:gauss_vs_synow}.). We find Equation \ref{eq:vs_vg} as the linear relationship between the (Ca II NIR3) HVF Gaussian velocities and those modeled by SYNOW (errors of the fitted parameters are shown in parentheses):

\begin{equation}
    v_{\rm HVF}^{\rm SYNOW} = 1.18(0.04) \cdot v_{\rm HVF}^{\rm Gauss} - 2212.29(838.38) ~{\rm km~s}^{-1}.
    \label{eq:vs_vg}
\end{equation}

Equation \ref{eq:vs_vg} is the observed relationship between the two velocity measurement methods, but it lacks information about the physical reason for the difference.

We believe that the reason for this systematic offset between the two sets of velocities is because the Doppler velocity calculated from the location of the absorption minimum via Gaussian fitting does not always reflect exactly the velocity of the line forming region itself. 

To demonstrate this, we created synthetic spectral lines of several different velocities with SYNOW, calculated their velocity with a Gaussian fit to the line minima and compared them to the true velocities (which is an input parameter in SYNOW, $v_{\rm HVF}$). 

\begin{figure}[ht]
    \centering
    \includegraphics[width=0.98\linewidth]{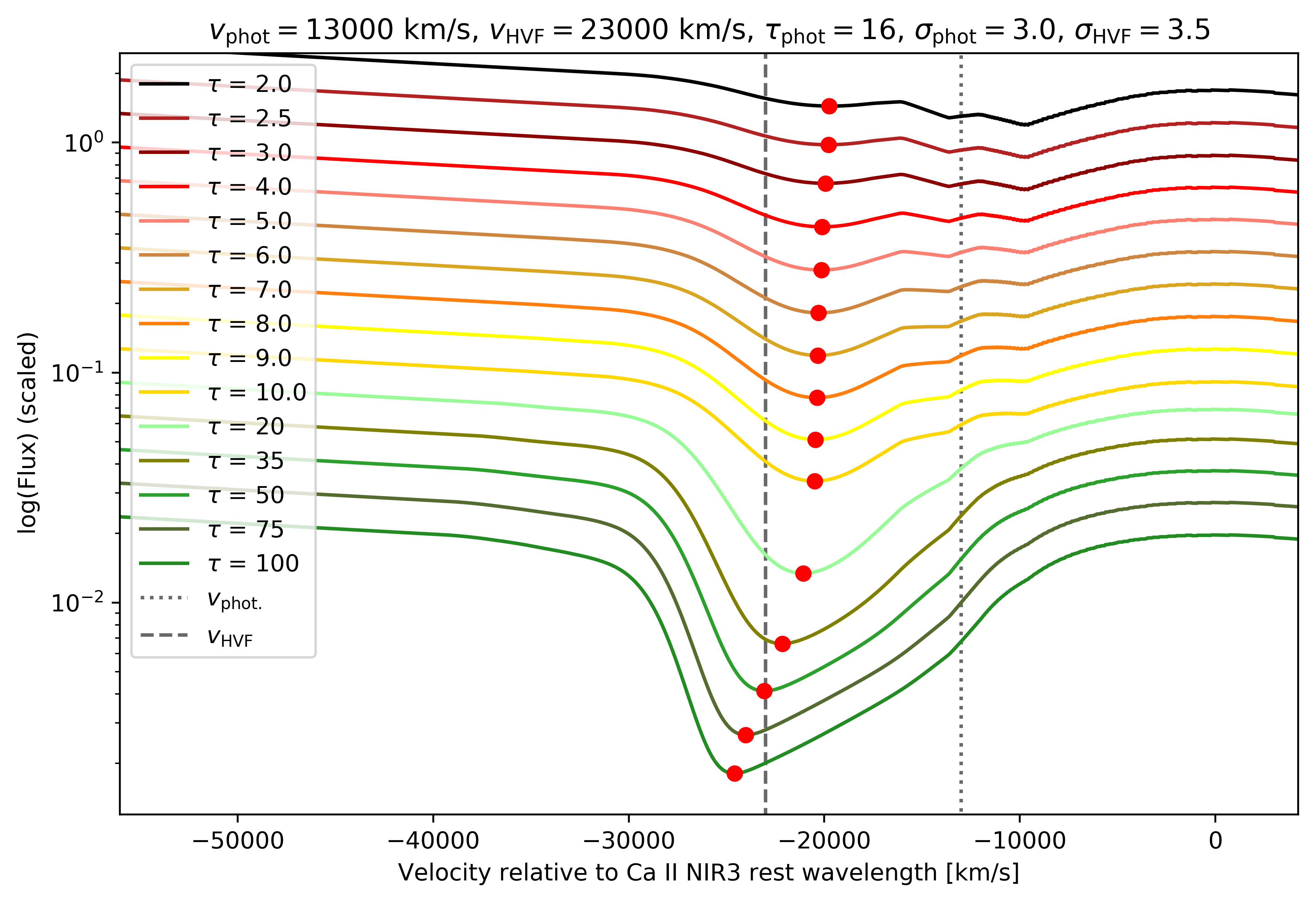}
    \caption{Shifting of the HVF line center with higher line optical depths, even when the velocity and other line parameters are unchanging. The different colored synthetic spectra show different line optical depths, the red markers show the HVF line minima. We mark the photospheric velocity with a vertical dotted line and the input HVF velocity with a vertical dashed line.}
    \label{fig:tau_delta_model}
\end{figure}

In Figure~\ref{fig:tau_delta_model} we show that when the optical depth increases, the line center shifts towards higher velocity regimes, even if the velocity of the line is kept constant. The line minimum is only consistent with the true input velocity when $\tau\approx 50$. In the $\tau$ range of the Ca II HVFs, this means that for all but the strongest of HVF lines, the velocities determined by finding the line minima (i.e. by fitting a  Gaussian) will underestimate the true velocity of the line forming region. For our sample, $\tau_{\rm HVF} \geq 50$ only occurs in 3 spectra, so in almost all cases, the Gaussian fits underestimate the modeled SYNOW velocities.

\begin{figure}[ht]
    \centering
    \includegraphics[width=0.98\linewidth]{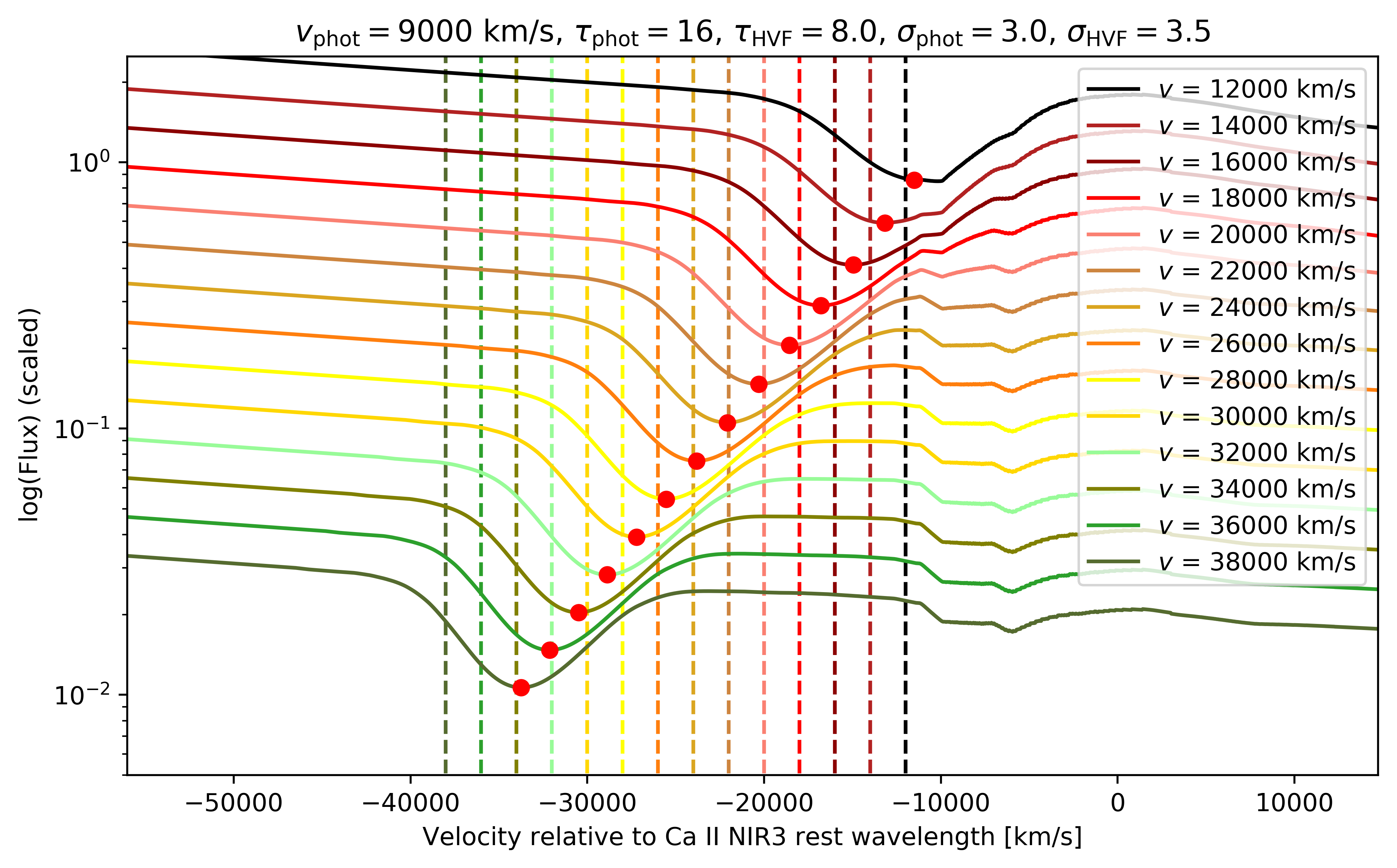}
    \caption{The different colored continuous lines show synthetic spectra with different HVF velocities. The corresponding vertical dashed line of the same color shows the velocity input into the SYNOW model to create the synthetic spectra. Red dots mark the HVF line minima. There is an increasing difference between the red dots and the corresponding vertical dashed lines with increasing input HVF velocities.}
    \label{fig:v_delta_model}
\end{figure}

A similar effect can be seen when we keep all line parameters constant and only increase $v_{\rm HVF}$. In this case, we find that the line minimum always lags behind the true line velocity, and this difference keeps getting larger with higher velocities (Fig. \ref{fig:v_delta_model}). For typical HVF velocities, the velocity of the line minimum always underestimates the true velocity of the line forming region. On Figure \ref{fig:v_delta_model}, we kept $\tau_{\rm HVF} = 8$, which is well under the limit where, because of the large line optical depth, the line minimum shifts to the left considerably ($\tau_{\rm HVF} \geq 50$ cases).

\begin{figure*}[ht]
    \centering
    \includegraphics[width=0.88\linewidth]{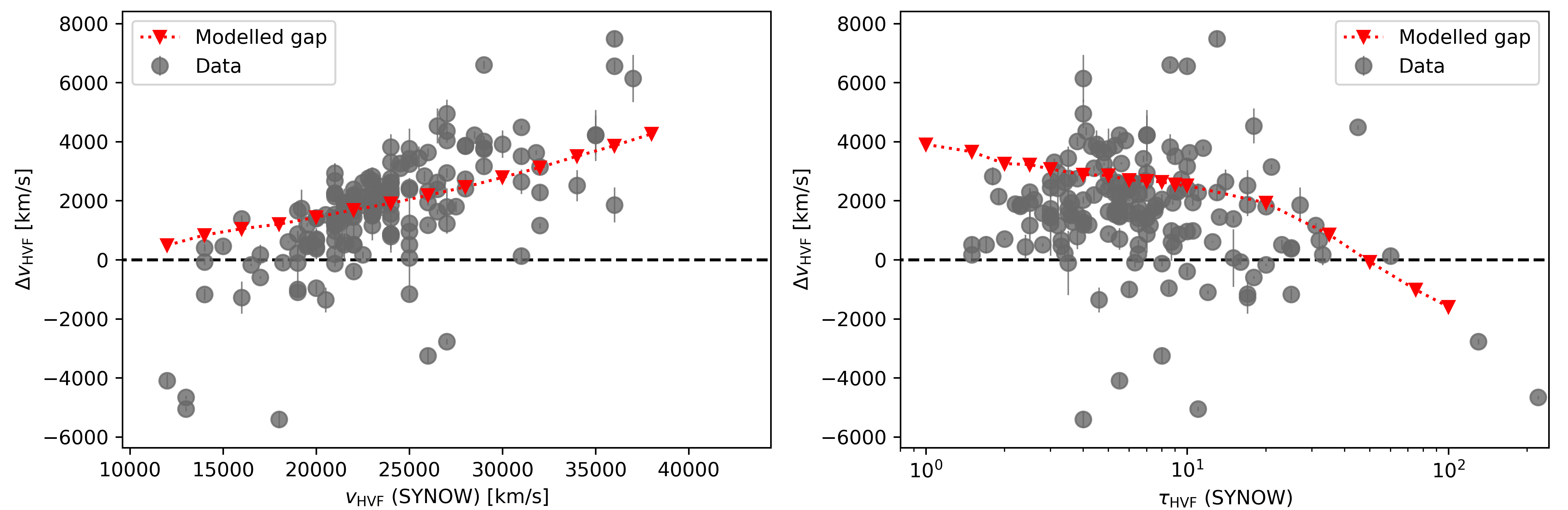}
    \caption{Grey markers show the differences between HVF velocities estimated with
    SYNOW and by Gaussian fitting to line profiles. The red triangles show the differences modeled in SYNOW between the input velocity and the velocity calculated from the position of the line minimum, based on Figures \ref{fig:tau_delta_model} and \ref{fig:v_delta_model}.}
    \label{fig:tau_v_gap_model}
\end{figure*}

If we compare these modeled differences to our observed $\Delta v_{\rm HVF}$ values, we find good agreement, as shown in Figure \ref{fig:tau_v_gap_model}, albeit with some scatter.

We can see that $\Delta v_{\rm HVF}$ is influenced by the velocity ($v_{\rm HVF}$) and the line optical depth ($\tau_{\rm HVF}$) of the HVF spectral line determined from SYNOW modeling.
To quantify this influence, we performed multiple variable linear regression on these parameters, and found the following relation between $\Delta v_{\rm HVF}$, $v_{\rm HVF}$ and $\tau_{\rm HVF}$:

\begin{equation}
\begin{split}
    \Delta v_{\rm HVF} = &-24.05(11.70) \cdot \tau_{\rm HVF} + 0.28(0.02) \cdot v_{\rm HVF} \\
    &-4336.48 (430.61).
    \label{eq:delta_v}
\end{split}
\end{equation}

Equation \ref{eq:vs_vg} shows simply the measured relationship between the velocities determined by the two methods, and does not include information about the physics behind the observed velocity difference. We show with Equation \ref{eq:delta_v} that the physical reason behind $\Delta v_{\rm HVF}$ is complex and due to multiple variables.

The Gaussian fitting of lines and spectral modeling are two common approaches used in the literature to determine the velocity of spectral lines, and in our case of HVFs. When comparing results from these two methods, it is important to keep in mind the difference in velocities caused by the behavior of spectral lines. Additionally, these features are inherently non-Gaussian, so fitting them with multiple Gaussian functions can lead to incorrect pEW estimations and feature identifications \citep{mulligan18}.

\subsection{Average velocities}
\label{sec:avg}
To characterize the overall (HVF or PVF) velocity of an event, it would be useful to calculate their phase averaged velocity.
However, for each SN, their phase of discovery, as well as the number and frequency of observations vary greatly. This will lead to certain events having only velocity measurements in the later (lower velocity) phases of their lives, while other SNe might only be observed in the early (high velocity) phases. Because of this phase-bias, we would inherently calculate a lower average velocity for those which were only observed in later times than those which were observed only in early phases, even if the velocity evolution of the two events would have been similar overall.

To work around this, for both methods of velocity measurement (Gaussian fitting and SYNOW modeling) and for all studied spectroscopic features (Si II PVF, HVF and Ca II PVF, HVF) we created average velocity curves. First, we manually multiplied the individual velocity curves with a value between 0.0 and 2.0 until the scatter of the entire sample was reduced. We then fitted the sample with an exponential function, thus creating an average velocity curve. By determining a scaling parameter to fit each individual SNe to this average curve, we obtained a uniformly sampled velocity function for all events. Finally, we computed the mean value of the fitted curve between phases -20 and 0 days for each SNe. We use these mean velocities in later sections as the ``average velocity'' to describe the general velocity of the SNe.

This method is useful for SNe Ia with regular velocity curves, but does not provide a good fit for a few peculiar events (SN~2021gtp or SN~2021J in our sample, for example). We find this drawback acceptable for the benefit of having a consistent method of calculating the average velocities for the whole sample.

\section{Results}
\label{sec:results}
In addition to our Gaussian fits and SYNOW modeling, we also collected parameters from the literature that define the light curve (SALT $x_1$, $\Delta m_{15}(\rm B)$) and subclass \citep{branch06,branch09,benetti05,wang09} of the supernovae in our sample. Table~\ref{tab:phys_data} shows these parameters together with some others inferred from SYNOW modeling: $v_{\rm SiII}$ at $t_{\rm max}$, the mean $R_{\rm HVF}$ for CaII and the mean velocity difference between the HVF and PVF ($\Delta v$ = $v_{\rm HVF} - v_{\rm PVF}$) for CaII.  

\begin{table*}[ht]
    \centering
    \tiny
    \setlength{\tabcolsep}{1.5pt}
    \begin{tabular}{lccccc|lccccc}
    \hline
    \hline
       SN  & $\Delta m_{15}(\rm B)$ & SALT $x_1$ & $v_{\rm SiII}(t_{\rm max})$ & $R_{\rm HVF}^{\rm avg.}$(CaII) & $\Delta v^{\rm avg.}$(CaII)  &    SN  & $\Delta m_{15}(\rm B)$ & SALT $x_1$ & $v_{\rm SiII}(t_{\rm max})$ & $R_{\rm HVF}^{\rm avg.}$(CaII) & $\Delta v^{\rm avg.}$(CaII)  \\
           &                        &            & (km~s$^{-1}$)        &  & (km~s$^{-1}$)  &       &                        &            & (km~s$^{-1}$)        &  & (km~s$^{-1}$) \\
       \hline
        SN~2009ig & 0.92 (0.04) $^e$ & 1.07 (0.07) $^i$ & 12926 (505)  &  3.37 (0.45) & 8693 (2866) &   SN~2013dy & 0.92 (0.01) $^d$ & 0.87 (0.02) $^i$ & 9459 (147)  &  2.40 (0.37) & 9405 (2583)  \\
        PTF~10bjs & 0.88 (0.03) $^c$ & -  & 12971 (151)  &  1.57 (0.42) & 8875 (7029)   &   SN~2017cbv & 0.99 (0.01) $^h$ & 0.62 (0.05) $^i$ & 7405 (1620)  &  4.47 (0.69) & 7116 (2740) \\
        PTF~10icb & 1.02 (0.05) $^c$ & 0.64 (0.40) $^i$ & 9704 (90)  &  -  & 9957 (411)     &   SN~2017hjy & 1.14 (0.01) $^g$ & -  & 7306 (73)  &  1.79 (0.19) & 7662 (1566) \\
        PTF~10qjl & 1.12 (0.04) $^c$ & -  & 10237 (73)  &  -  & 9770 (689)  &   SN~2017ln & -  & -  & 11077 (60)  &  2.03 (0.28) & 8151 (1846) \\
        PTF~10ygu & 0.90 (0.03) $^c$ & -  & 10770 (445)  &  -  & 7307 (1494)    &  SN~2018gv & -  & 0.59 (0.04) $^i$ & 10281 (79)  &  4.74 (0.21) & 8319 (614)  \\
        SN~2010ai & 1.34 (0.01) $^c$ & -1.65 (0.09) $^i$ & 10606 (199)  &  0.14 (0.44) & 7944 (5284)    &   SN~2018ilu & -  & -  & 9628 (856)  &  1.58 (0.16) & 7258 (1948)   \\
        SN~2010dm & 0.80 (0.01) $^c$ & -  & 10131 (369)  &  -  & 9606 (628)     & SN~2018oh & 0.99 (0.01) $^g$ & 0.88 (0.01) $^h$ & 9806 (123)  &  1.34 (0.13) & 9372 (764) \\
        SN~2010it & -  & -  & 12327 (404)  &  2.40 (0.23) & 9823 (3007)     &  SN~2020aabz & -  & -  & 10131 (89)  &  1.64 (0.23) & 8166 (1440)  \\
        SN~2010iw & -  & -  & 10054 (353)  &  2.13 (0.29) & 8665 (2874)     &   SN~2021J & 1.03 (0.05) $^j$ & -  & 11251 (270)  &  1.76 (0.20) & 9277 (6472)   \\
        SN~2010ke & -  & -  & 6507 (1179)  &  -  & 9097 (1476)  &    SN~2021gtp & -  & -0.19 (0.11) $^k$ & 10770 (308)  &  3.11 (1.18) & 7343 (5216) \\
        SN~2010kg & 1.37 (0.06) $^c$ & -1.29 (0.19) $^i$ & 8807 (565)  &  3.25 (0.42) & 11141 (2608)    &  SN~2021hiz & -  & -0.46 (0.04) $^k$ & 9598 (139)  &  0.70 (0.10) & 7670 (996) \\
        SN~2011ao & -  & 0.42 (0.14) $^i$ & 10645 (636)  &  1.59 (0.11) & 9045 (4356)   &  SN~2022erw & -  & -0.13 (0.16) $^n$ & 12251 (345)  &  2.22 (0.60) & 7269 (3913) \\
        SN~2011by & 1.14 (0.03) $^a$ & -0.33 (0.04) $^i$ & 9491 (113)  &  -  & 8445 (326)   &  SN~2022hkc & -  & -  & 11409 (0)  &  -  & 9347 (116)  \\
        SN~2011dm & -  & -  & 12795 (93)  &  -  & 8560 (481)    &  SN~2022yhl & -  & -  & 14525  (101)  &  5.64 (0.41) & 7638 (3207)   \\
        SN~2011fe & 1.18 (0.01) $^g$ & -0.55 (0.13) $^i$ & 10053 (56)  &  1.30 (0.20) & 7937 (3582)     & SN~2022zut & -  & 0.05 (0.08) $^k$ & 12262 (97)  &  1.70 (0.07) & 6567 (365)  \\
        SN~2011fg & -  & -  & 9385 (263)  &  -  & 9139 (1474)   & SN~2023alb & -  & -  & 10022 (825)  &  1.16 (0.47) & 2830 (356)  \\
        SN~2011hb & -  & 0.15 (0.36) $^i$ & 10372 (686)  &  2.04 (0.14) & 10331 (1042)  &   SN~2023bee & 0.79 (0.00) $^l$ & 1.36 (0.03) $^k$ & 11091 (695)  &  2.66 (0.52) & 6384 (4736)    \\
        SN~2012G & -  & -  & 9172 (401)  &  -  & 2493 (1146)    &  SN~2023eoc & -  & -  & 11456 (456)  &  4.76 (1.09) & 7229 (5987)  \\
        SN~2012bh & -  & -0.41 (0.26) $^i$ & 9704 (45)  &  -  & 9224 (759)  &  SN~2023ktw & -  & -  & 11487 (225)  &  3.56 (0.87) & 5048 (3333) \\
        SN~2012bm & -  & -  & 9826 (114)  &  2.31 (0.27) & 12669 (1055)     &   SN~2023pbe & -  & -1.38 (0.04) $^n$ & 10708 (366)  &  0.62 (0.56) & 6510 (4775)  \\
        SN~2012cg & 0.83 (0.05) $^c$ & 0.49 (0.02) $^i$ & 10027 (404)  &  2.19 (0.19) & 8218 (2342)     &  SN~2023tky & -  & -0.00 (0.18) $^n$ & 10516 (101)  &  0.85 (0.30) & 8893 (1783)  \\
        SN~2012da & -  & -  & 10260 (281)  &  0.91 (0.12) & 8189 (763)  &   SN~2023vyl & -  & 0.19 (0.04) $^n$ & 10344 (864)  &  1.02 (0.29) & 8089 (3060)  \\
        SN~2012fr & 0.85 (0.05) $^c$ & 1.16 (0.04) $^i$ & 7037 (426)  &  8.73 (2.59) & 11588 (1020)     &  SN~2023xtf & -  & -  & 8426 (723)  &  1.39 (0.58) & 7284 (370)\\
        SN~2012hj & -  & -  & 8699 (1176)  &  7.44 (0.57) & 10816 (1088)    &   SN~2023xuz & -  & -  & 8639 (1076)  &  4.87 (2.54) & 7649 (1701)   \\
        SN~2012ht & 1.39 (0.05) $^b$ & -1.16 (0.04) $^i$ & 10587 (82)  &  0.43 (0.36) & 7398 (3366)     &   SN~2023zvn & -  & -  & 11644 (106)  &  2.22 (0.42) & 7733 (2044) \\
        SN~2013ag & -  & -  & 11729 (62)  &  -  & 6479 (851)    &  SN~2024gy & 1.12 (0.00) $^m$ & 0.50 (0.08) $^n$ & 8902 (346)  &  3.27 (0.27) & 11954 (2501)\\
        SN~2013be & -  & -  & 11409 (221)  &  -  & 8077 (917)   &   SN~2024iei & -  & -  & 9820 (297)  &  1.74 (0.33) & 8623 (3220)  \\
        SN~2013di & -  & -  & 9811 (76)  &  -  & 8952 (242)     &   SN~2024xyn & -  & -  & 10894 (843)  &  2.07 (0.24) & 8868 (2902)\\
        
    \hline    
    \end{tabular}
    \caption{Light-curve decline rate, stretch parameter, Si II photospheric velocity at maximum brightness, average HVF strength and HVF-PVF separation for the Ca II NIR triplet of each supernova in our sample. \\a: \cite{silverman13}, b: \cite{yamanaka14}, c: \cite{silverman15}, d: \cite{pan15}, e: \cite{chakradhari19}, f: \cite{dimitriadis19}, g: \cite{ktr20}, h: \cite{wang20}, i:  \cite{brout22}, j: \cite{gallegocano22}, k: \cite{bora24}, l:  \cite{wang24}, m: \cite{li26}, n: this paper.}
    \label{tab:phys_data}
\end{table*}

\begin{figure}[ht]
    \centering
    \includegraphics[width=0.96\linewidth]{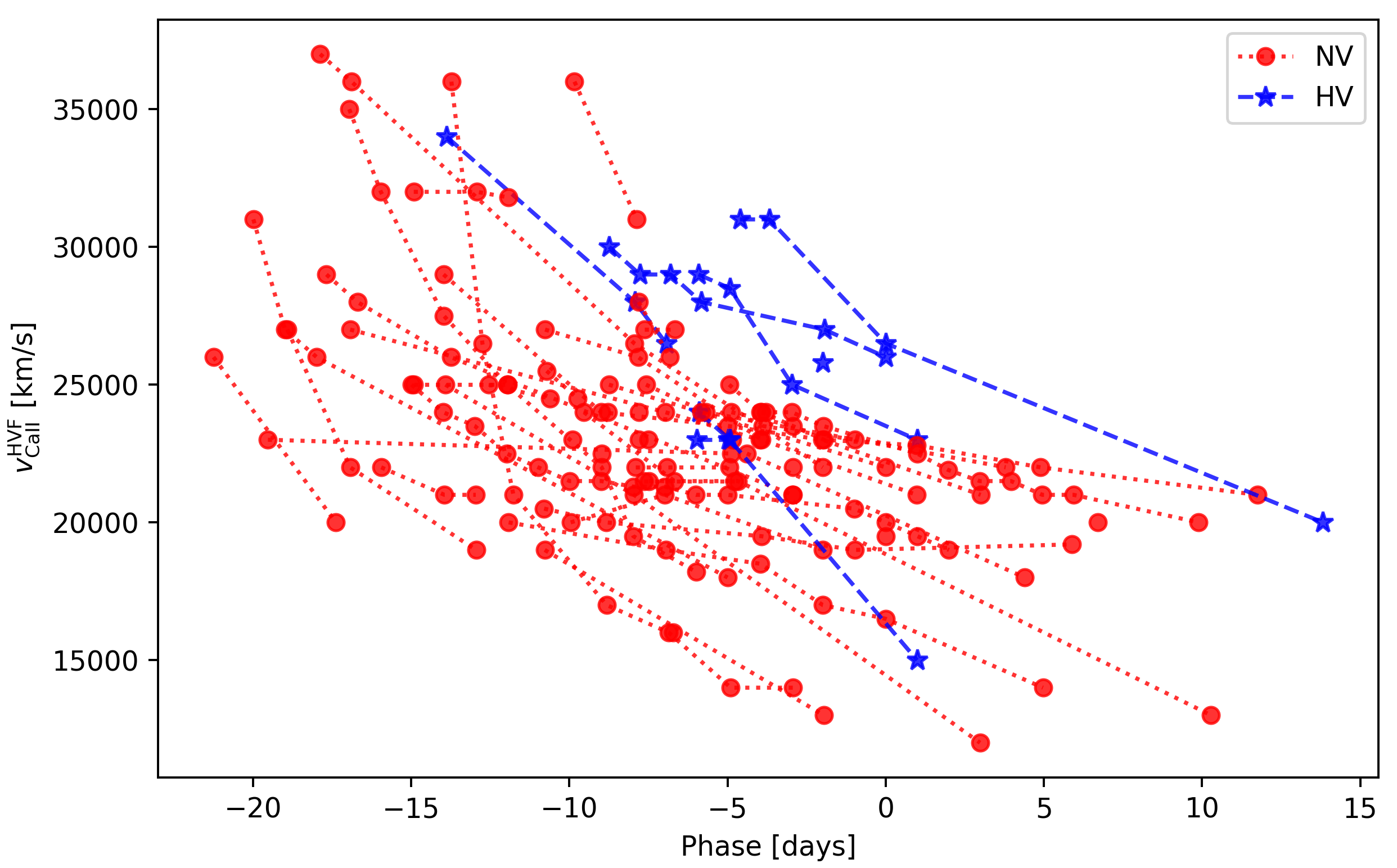}
    \caption{Ca II HVF velocities (modeled with SYNOW) against phase, with the measurements of each supernova being connected. The HV supernovae are denoted with blue, the NVs with red. Wang HV events tend to have higher Ca II HVF velocities.}
    \label{fig:wang_HVF}
\end{figure}

First we use the $v_{\rm Si}^{\rm tmax}$ values fitted in Section \ref{sec:gauss} to determine the \cite{wang09} classification of individual events in our sample. We find that almost all objects in our sample are members of the normal velocity (NV) group, with only 7 having high enough photospheric velocities to be considered high velocity (HV) supernovae (Figure \ref{fig:wang_HVF}, Table \ref{tab:sample2}).  
These HV supernovae also show high Ca II HVF velocities compared to the rest of the NV sample.

The low number of HV objects among our sample of SNe is consistent with the conclusion of several earlier studies \citep{childress14, maguire14, silverman15} that HV SNe Ia show only weak HVFs.  

\subsection{Connections with the light curve stretch parameter}
\label{sec:x1_correlations}

\begin{figure*}[ht]
    \centering
    \includegraphics[width=0.88\linewidth]{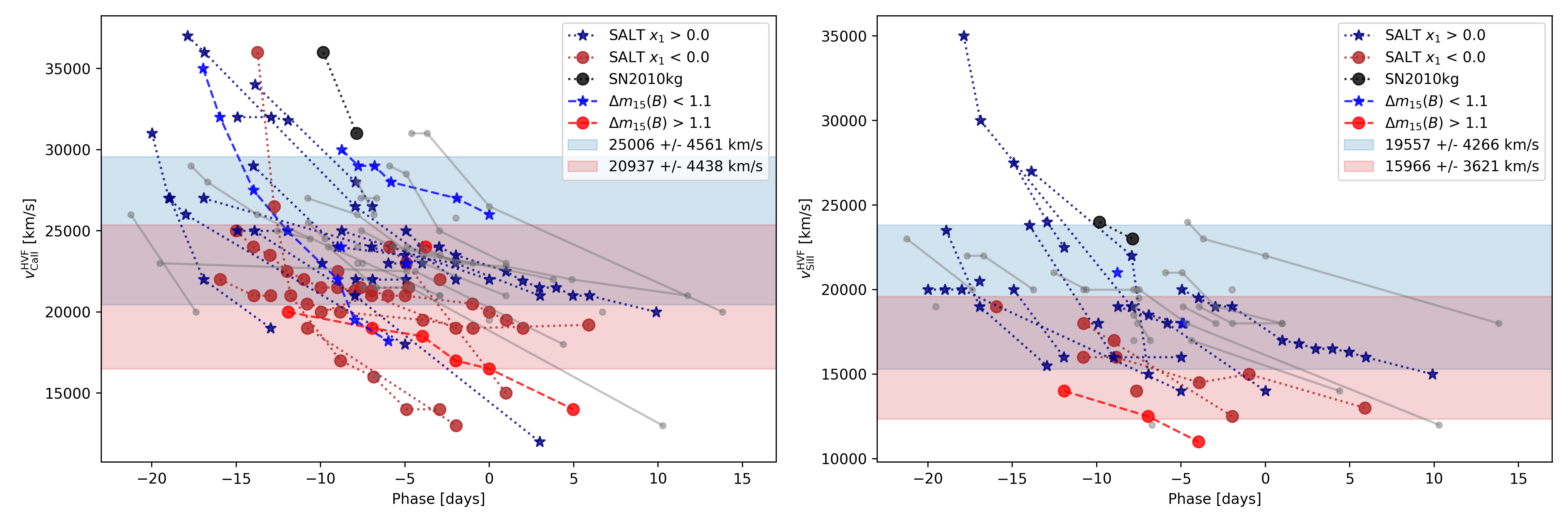}
    \caption{HVF velocities of Ca II (left) and Si II (right) from SYNOW modeling against phase. Slow-decliners are shown in blue, fast decliners in red, and those without this information in grey. The blue and red bands show the average ($\pm$ one standard deviation) velocity of their respective groups. The outlying fast decliner with high HVF velocities, but small SALT $x_1$ is the anomalous SN~2010kg, marked with black circles.}
    \label{fig:m15B_vHVF}
\end{figure*}

The $\Delta m_{15} (\rm B)$ and SALT $x_1$ parameters describe the shape of the light curve that correlates with the peak brightness through the Phillips relation.
Similar to previous studies \citep{childress14, maguire14, silverman15, zhao15}, we also find that slow-decliner (bright) SNe tend to have higher HVF velocities for both Ca II and Si II than fainter events (Figure \ref{fig:m15B_vHVF}.). Brighter supernovae (with $\Delta m_{15}(B) < 1.1 \text{ mag}$ or SALT $x_1 > 0.0$) have an average Ca II HVF velocity of $\sim 25000$ km~s$^{-1}$, while fainter objects (with $\Delta m_{15}(B) > 1.1 \text{ mag}$ or SALT $x_1 < 0.0$) show an average of $v_{\rm Ca II}^{\rm HVF} \approx 21000$ km~s$^{-1}$. The Si II HVFs show the same behavior but at lower velocities: $\sim 20000$ km~s$^{-1}$ for bright events, and $\sim 16000$ km~s$^{-1}$ for faint ones. 

In Figure \ref{fig:m15B_vHVF} (and also in later sections), SN~2010kg is an outlier:  even though it is characterized as a fast decliner by both SALT $x_1$ and $\Delta m_{15}(B)$, it has high HVF velocities. In the earliest spectra, the Ca II NIR3 HVF shows such a high velocity that the absorption feature blends together with the neighboring O I $\lambda7773$\AA\ line. This blending might cause some overestimation of the early Ca II HVF velocity, although it does not explain why the Si II $\lambda 6355$\AA\ HVF also shows higher velocity than we would expect from the LC decline rate.

From now on, to uniformly characterize the decline rate for the whole sample, we use only the SALT $x_1$ parameter, as it was more frequently available for our sample SNe than $\Delta m_{15} (\rm B)$.

\begin{figure*}[ht]
    \centering
    \includegraphics[width=0.88\linewidth]{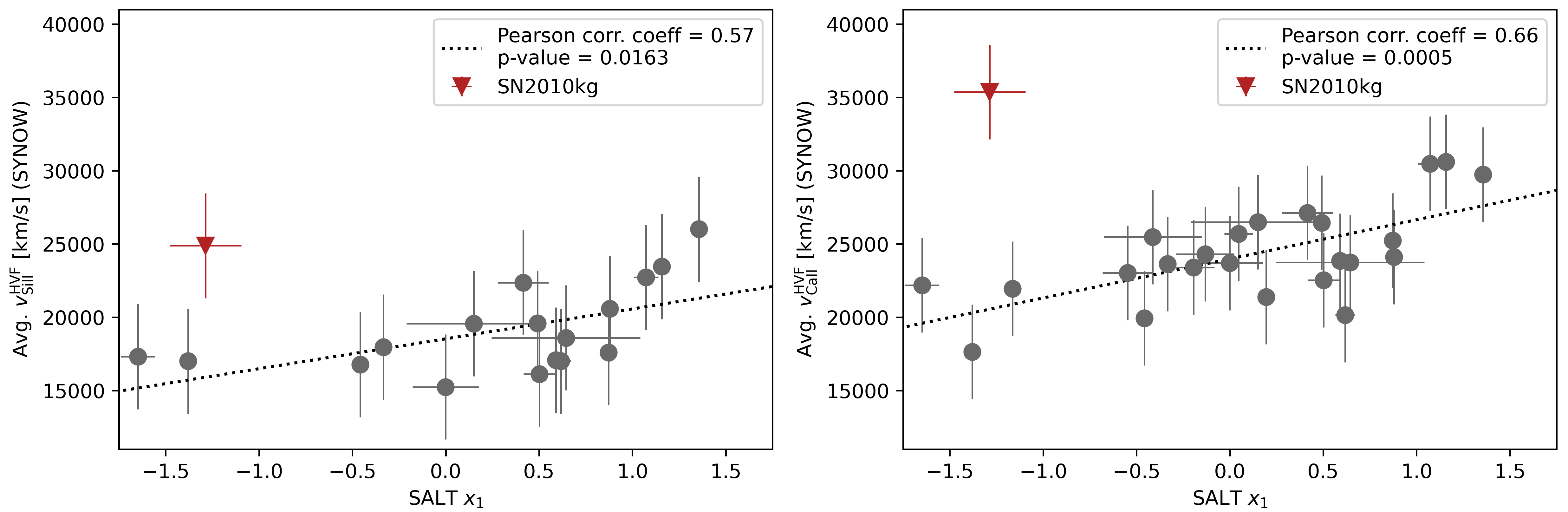}
    \caption{Relations of the average HVF velocity for an object as modeled by SYNOW against its light-curve width (characterized by the SALT $x_1$ parameter). The left panel shows the result for the Si II $\lambda6355$ line, the right panel for the Ca II NIR triplet. The outlier SN~2010kg is marked with red triangles in both plots, and was excluded from statistical analyses.}
    \label{fig:x1_mosaic}
\end{figure*}

On Figure \ref{fig:x1_mosaic}, we show the relationships between the average SYNOW modeled velocity (see Sec \ref{sec:avg}.) of the Si II  and Ca II HVF lines and the decline rate of a SN Ia (described by SALT $x_1$). For the Si II and Ca II HVFs, we find a positive correlation between the average velocities and the SALT $x_1$ parameter of the SN, although for the Si II HVF, the sample is limited. 
In the case of the Ca II PVF, we did see a weak tendency for the average velocity to increase with $x_1$, but we did not find this to be statistically significant, and for this reason it was not included in the final figure.

When repeating this analysis with the Gaussian average velocities, we find similar positive correlations between SALT $x_1$ and the Si II HVF, Ca II HVF and Ca II PVF as well. We find no correlation between the photospheric Si II velocity and SALT $x_1$.

To summarize, across both methods of velocity measurement we find a similar positive correlation between SALT $x_1$ and the Si II and Ca II HVF phase-averaged velocities. A correlation between the photospheric Ca II and $x_1$ can be seen in the case of the Gaussian method, but this same relation is not clear in the case of the SYNOW velocities. The Si II photospheric velocities show only scattering and no correlation for both methods.

\cite{hakobyan26} also did not find a significant correlation between the photospheric Si II velocity around maximum time and the decline rate $\Delta m_{15}$ for the entire sample of \cite{silverman15}. However, when they excluded HVF-strong events, and only considered $R_{\rm HVF} \leq 0.5$, a strong negative correlation emerged. They suggest that strong HVFs might modulate the velocity structure of the ejecta and mask the relationship between ejecta velocity ($v_{\rm Si II}$) and explosion energetics ($\Delta m_{15}$/SALT $x_1$).

\begin{figure}[ht]
    \centering
    \includegraphics[width=0.95\linewidth]{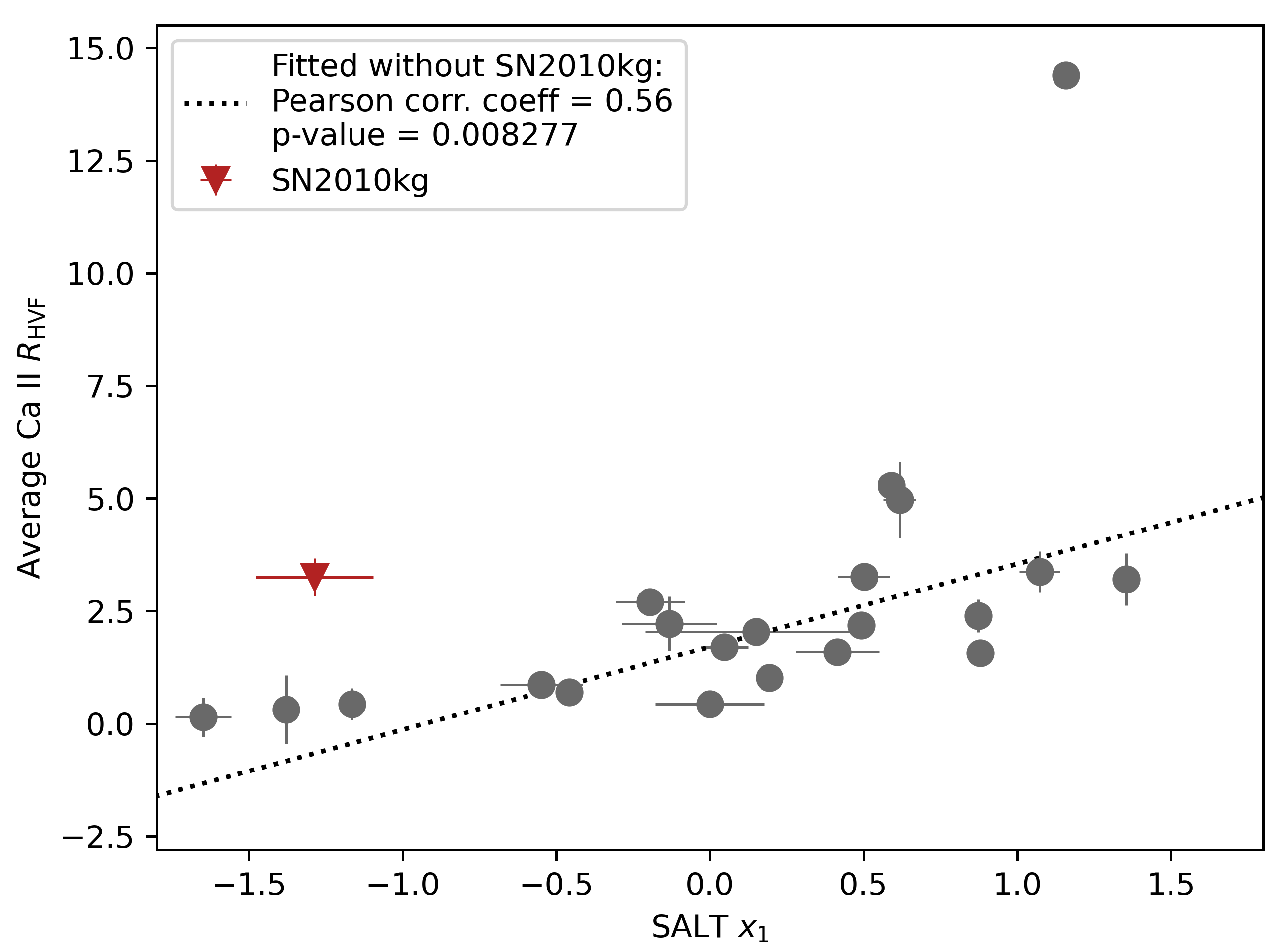}
    \caption{Average Ca II HVF strength of each object compared to their light-curve stretch, measured with the SALT $x_1$ parameter.}
    \label{fig:x1_R}
\end{figure}

When comparing the average Ca II HVF strength and SALT $x_1$, we find a significant positive correlation between the two, showing that brighter events also display stronger HVFs (Fig \ref{fig:x1_R}.). 

\begin{figure}[ht]
    \centering
    \includegraphics[width=0.95\linewidth]{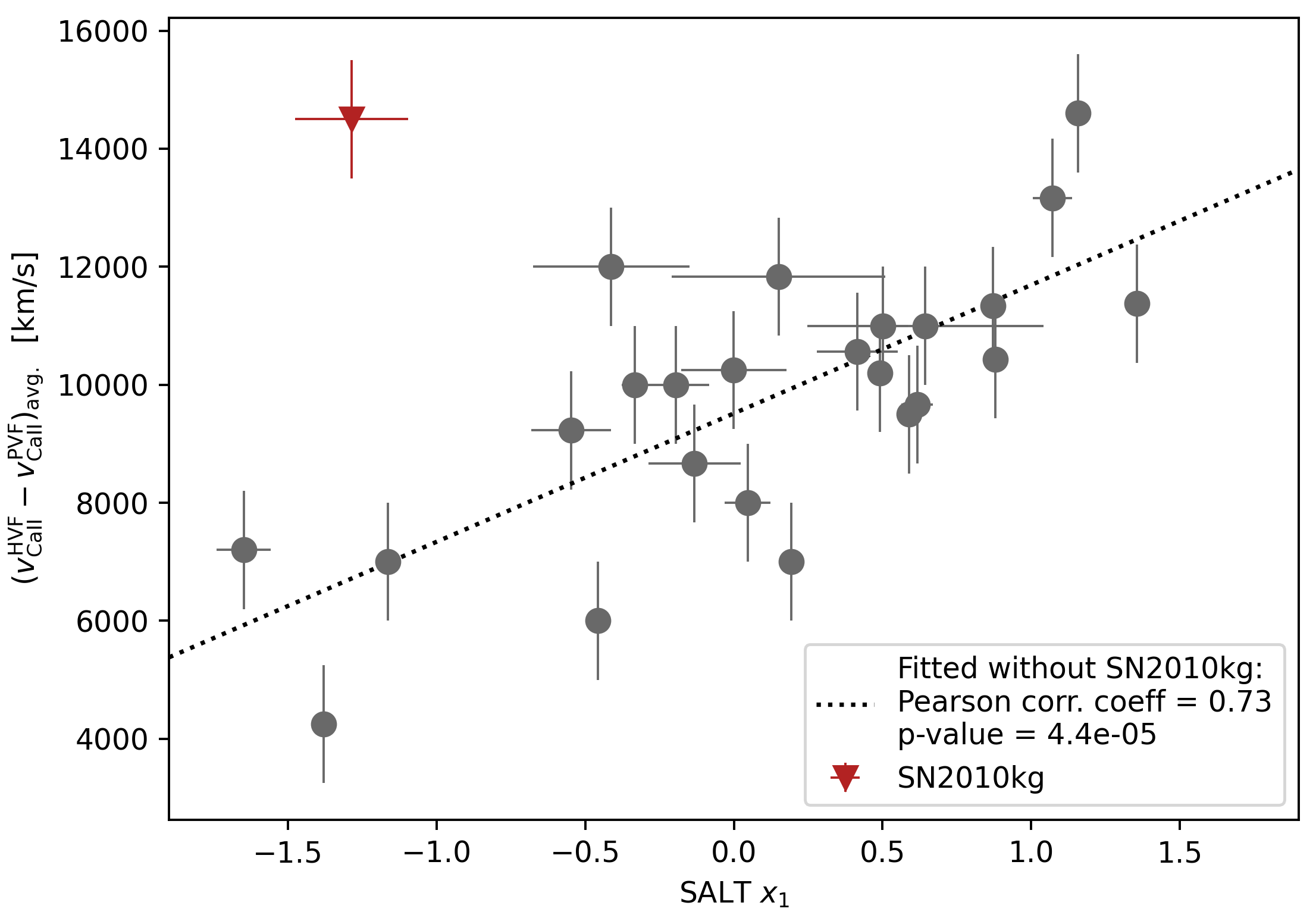}
    \caption{Average separation between the SYNOW modeled Ca II HVF and PVF of each object compared to their light-curve stretch parameter, measured with the SALT $x_1$ parameter.}
    \label{fig:x1_gap}
\end{figure}

While the average velocity difference between the HVF and PVF components remains mostly unchanged for all phases, Figure~\ref{fig:x1_gap} shows that it does grow with SALT $x_1$, meaning that slow-decliner (brighter) supernovae show a bigger separation between their high-velocity and photospheric layers.

This is in agreement with our other results. Brighter, slow-declining SNe Ia tend to have higher velocity and stronger HVFs (Figures \ref{fig:x1_mosaic}, \ref{fig:x1_R}). The larger separation with increasing $x_1$ can come from a higher HVF velocity and/or a lower PVF velocity. We can see on Figure \ref{fig:x1_mosaic} that while the Ca II HVF strongly correlates with the light-curve stretch, the Ca II PVF does not, so it is to be expected that the separation between the two components would increase with SALT $x_1$. In the case of the Gaussian velocities, where we do see a correlation between the Ca II PVF and $x_1$, the increase of the PVF velocity with $x_1$ is less steep than that of the Ca II HVF, again allowing for the separation between the two to grow with increasing stretch. 

\begin{figure}[ht]
    \centering
    \includegraphics[width=0.96\linewidth]{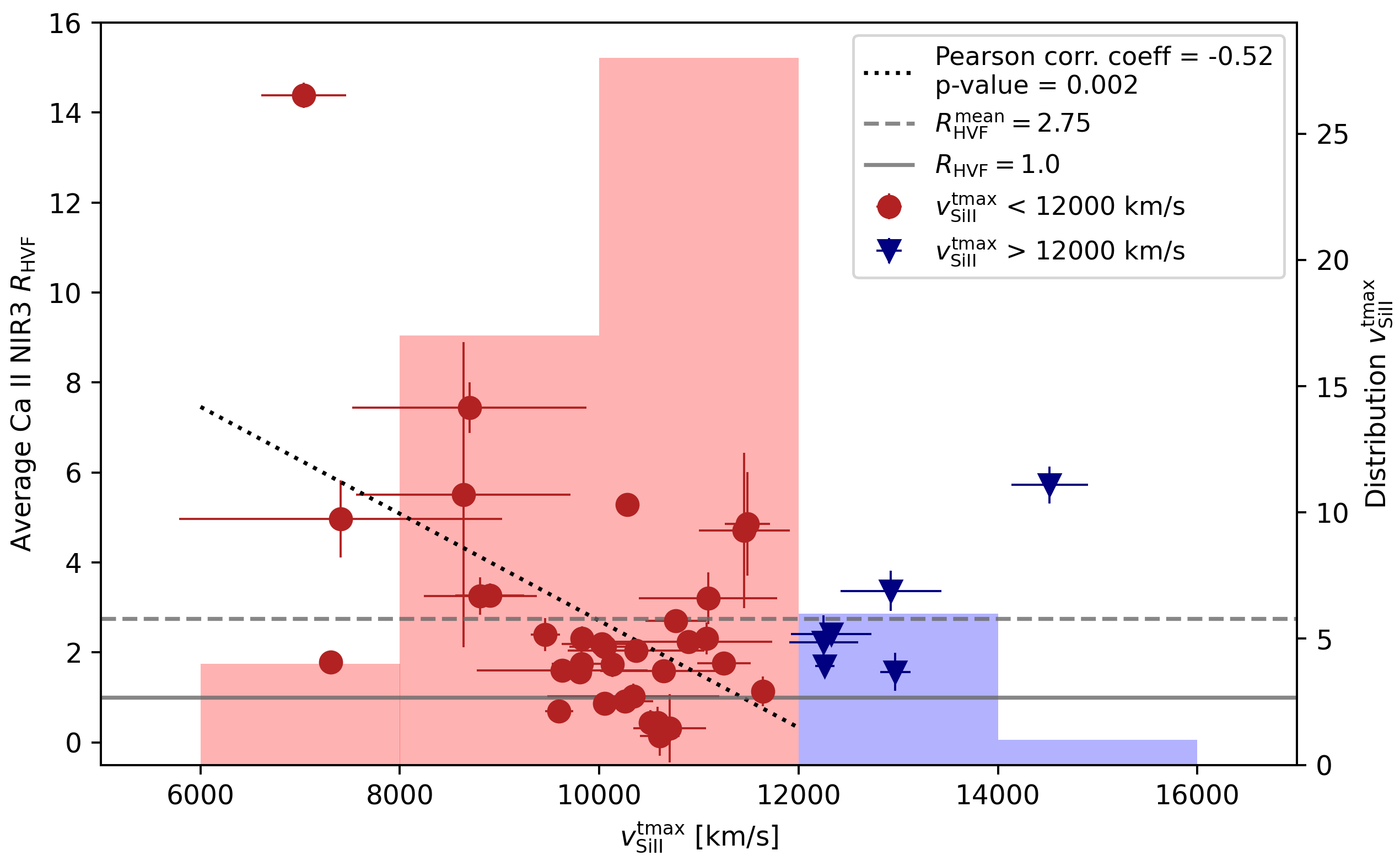}
    \caption{Average Ca II HVF strength of each object compared to their Si II $\lambda6355$ photospheric velocity at the time of maximum light. With red, we mark the \cite{wang09} NV, with blue the HV objects. The solid grey line shows $R_{\rm HVF} = 1.0$, above which an HVF is considered to be strong compared to its photospheric component. The grey dashed line marks the mean HVF strength. The black dotted line shows the linear fit done to the NV objects. The histograms in the background show the distribution of the Si II velocities for both groups. For the NV events, we find a tendency for $R_{\rm HVF}$ to decrease with increasing $v_{\rm Si II}^{\rm tmax}$, spanning a wide range, while the HV events tend to show $R_{\rm HVF}$s around the mean.}
    \label{fig:si_RCa}
\end{figure}

There also seems to be a connection between the Ca II NIR3 HVF strengths and the Si II photospheric velocity at maximum ($v_{\rm Si II}^{\rm tmax}$). SNe with the strongest Ca II (and Si II) HVFs are almost all Wang NV events, and the HVF strength seems to decrease with increasing $v_{\rm Si II}^{\rm tmax}$, covering a large range of $R_{\rm HVF}$s (see Fig \ref{fig:si_RCa}.). Wang HV objects do not seem to follow this trend, and their HVF strengths tend to cluster around the mean $R_{\rm HVF}$ value. However, HV events only make up  $12.5 \%$ of our sample, which is not only low compared to recent statistics \citep{burgaz25}, but also makes it difficult to draw conclusions about the subclass.

As both the average strength and the velocity of the Ca II HVF seems to positively correlate with the light curve stretch parameter $x_1$, we can expect a relationship between the strength and the velocity parameters as well.
Indeed, for the whole sample, we find a positive trend (with $p = 0.0076$) between the average velocity and the average strength of a Ca II HVF, meaning that for fast HVFs, their features will also be on average stronger than their photospheric components.

Overall, in agreement with previous findings, we conclude that brighter, slow-declining SNe Ia generally have faster, stronger HVFs, bigger separation between their high-velocity and photospheric regions, and SNe Ia with strong HVFs tend to be Wang NV types based on their Si II velocity near maximum light.

SALT $x_1$ and other light-curve shape or decline rate parameters primarily describe the photometric evolution of SNe Ia. In this section, we showed that the spectroscopic evolution of HVFs are also strongly connected to the light-curve shape, despite not leaving any obvious signs of their presence or strength on the light curves themselves.

\subsection{Si II HVFs}
\label{sec:SiII_HVFs}

HVFs of Si II were first confirmed by \cite{mazzali05b} and have been found in many early spectra since then, but they are still less frequently observed than Ca II HVFs.
According to \cite{harvey25}, HVFs in the Si II $\lambda$6355 line appear in $\approx 76 \%$ of Type Ia supernova spectra $t < -11$ days before maximum (while according to \cite{maguire14}, HVFs of Ca II NIR3 appear in $\approx 95\%$ of cases at $t<-5$ days), but quickly fade, and their percentage drops to about 29\% at $t < -6$ days before maximum. Thus, their detection is more difficult than that of Ca II, which is most likely the reason why only 40 out of our 56 objects ($\approx 71$ \%) display a visible Si II HVF. Out of all early spectra ($t<-11$ days), we find $64\%$ of them show a Si II HVF, which is a lower percentage than what \cite{harvey25} found.

In most cases, the Si II HVF blends together with the photospheric component and the only sign of its presence is a change in the shape of the Si II line. Because of this, determining their velocity, or the phase at which they disappear, is not trivial. This might be the cause of the previously stated difference between our percentage and that of \cite{harvey25}.

\begin{figure*}[ht]
    \centering
    \includegraphics[width=0.85\linewidth]{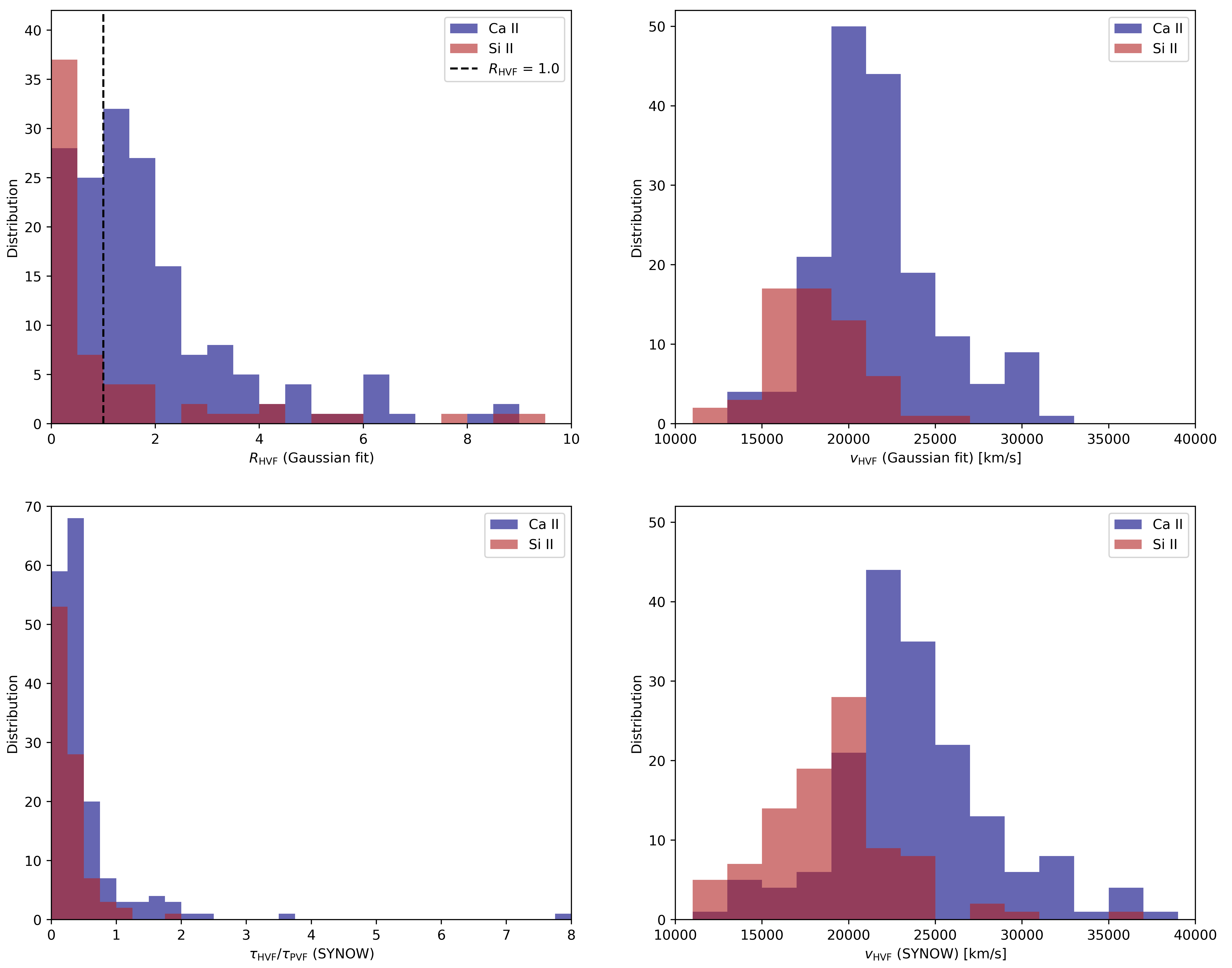}
    \caption{Top: Histograms of the strength (left) and velocity (right) distributions of the Si II $\lambda6355$ (red) and Ca II NIR triplet (blue) HVFs for all spectra and phases in our sample from Gaussian fittings. The black dashed line on the left hand plot shows $R_{\rm HVF} = 1.0$.\\ Bottom: Histograms of the ratio of HVF and PVF line optical depths (left) and HVF velocity (right) for the Si II (red) and Ca II (blue) HVFs obtained from SYNOW modeling.}
    \label{fig:si_ca_hist}
\end{figure*}

Based on our sample, we find that HVFs of the Si II $\lambda6355$ line are generally slower, weaker, less common, and fade faster than those of the Ca II NIR triplet. The comparison between the strength ($R_{\rm HVF}$) and velocity ($v_{\rm HVF}$) of the Si II and Ca II HVFs is shown on the top row of Figure \ref{fig:si_ca_hist}. We can see that the strength of the Si II HVFs rarely exceed $R_{\rm HVF} = 1.0$, which means that most often they are equal to, or even weaker in strength than their photospheric components, which is another reason why their detection is difficult.  Velocity-wise, they are generally around $\approx 5000$ km~s$^{-1}$ slower than the HVFs of Ca II, hinting at a deeper location within the ejecta.

\begin{figure}[ht]
    \centering
    \includegraphics[width=0.93\linewidth]{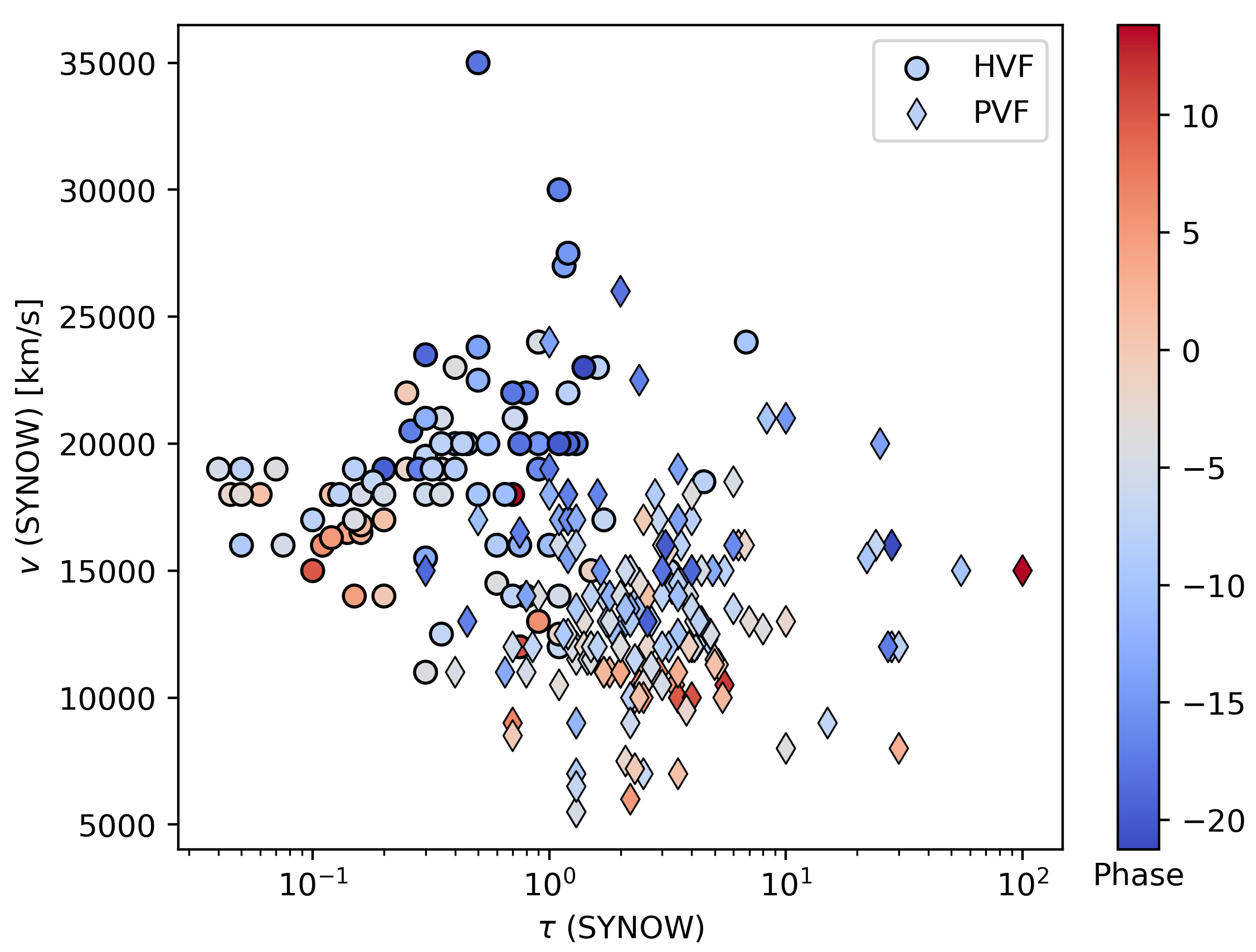}
    \caption{Velocity versus line optical depth for the Si II components according to the SYNOW modeling. HVFs are shown with thick bordered circles, PVFs with thin diamonds. Coloring is according to phase.}
    \label{fig:Si_tau_v}
\end{figure}

This trend can be seen in the results of the SYNOW modeling as well: the Si II HVF has lower velocities than the Ca II, and the ratio of $\tau_{\rm HVF}$ to $\tau_{\rm PVF}$ (which we use as a proxy for $R_{\rm HVF}$) is also smaller compared to that of the Ca II (bottow row of Fig \ref{fig:si_ca_hist}).
The SYNOW modeling also shows that the line optical depths and velocities (Fig. \ref{fig:Si_tau_v}.) of the Si II HVFs are lower compared to the Ca II HVFs (see Fig \ref{fig:Ca_tau_v}), and as such, they are closer to their photospheric components both in velocity and line optical depth. In Figure \ref{fig:Si_tau_v}, we also see a tendency for the Si II PVFs to have higher line optical depths ($\tau \geq 10$) than most of the HVFs, once again demonstrating the weakness of Si II HVFs. 

The difference between Si II and Ca II HVFs is most likely due to differences in the excitation energies of the Si II $\lambda 6355$ \AA~ and Ca II NIR3 lines (8.12 vs. 1.7 eVs, respectively). According to \cite{zhao15}, the Ca II triplet is easily formed in the outer (higher velocity), lower temperature regions of the expanding ejecta, while the Si II line is more sensitive to temperature. The abundance of Ca and Si might be the same at all velocities of the ejecta, but because of their temperature dependence, Si II lines can only form in the deeper, hotter regions, which is reflected in their lower velocities.
This might also be the reason why Si II HVFs tend to fade faster than the Ca II. As the ejecta expand and cool, the outer HVF region cools sooner than the main ejecta and cannot provide the needed environment for Si II lines to form at velocities typical of HVFs.

\subsection{Radius evolution}
\label{sec:t_radius}

\begin{figure}[ht]
    \centering
    \includegraphics[width=0.93\linewidth]{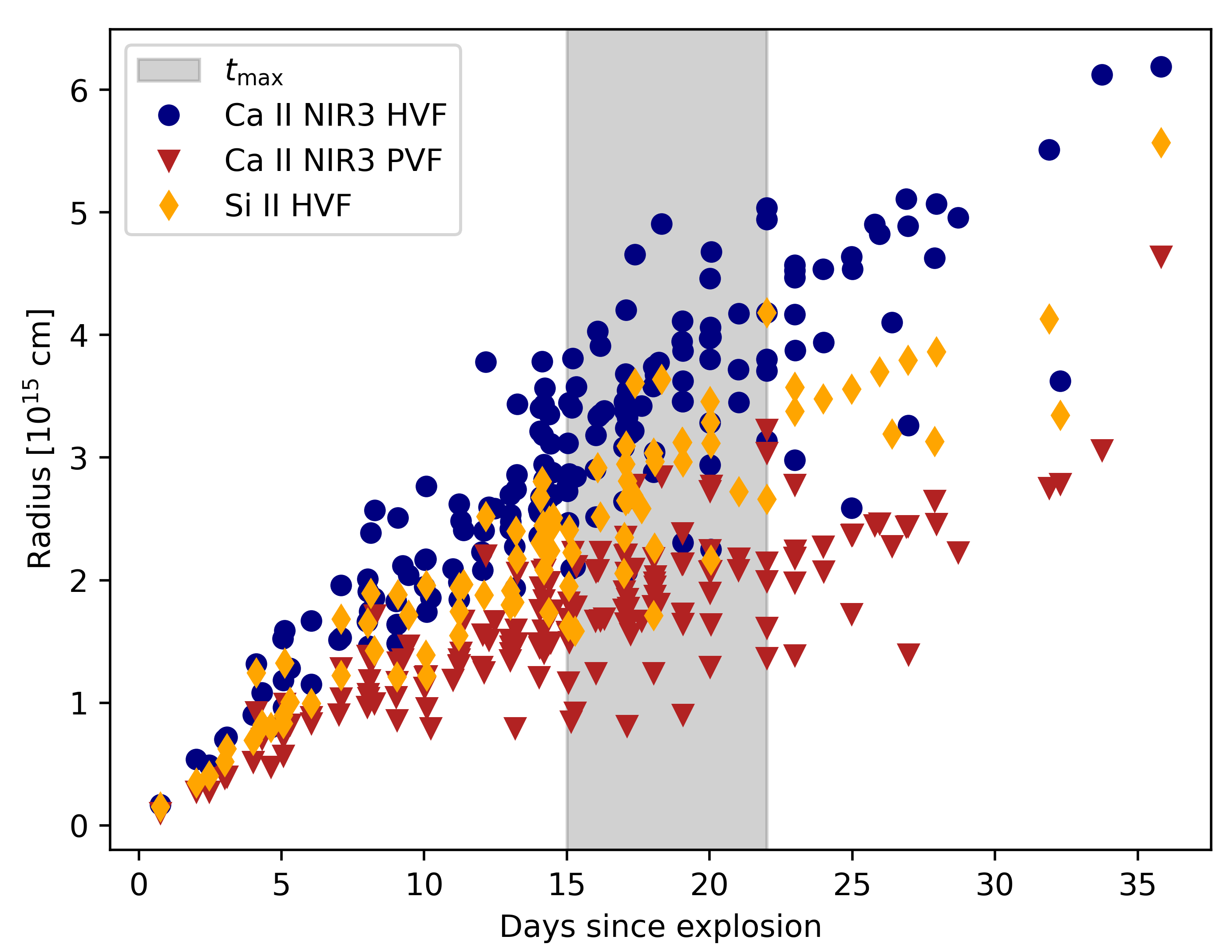}
    \caption{Distance traveled by the ejecta assuming homologous expansion, as traced by the SYNOW modeled velocities of different spectral features. Red triangles mark the photospheric Ca II NIR3 line, yellow diamonds the Si II $\lambda$6355 \AA~ HVF, and blue circles the Ca II NIR3 HVFs. The gray band highlights the range for the time of maximum light.}
    \label{fig:t_r}
\end{figure}

If we assume homologous expansion of the ejecta after explosion, we can convert the SYNOW velocity curves into radius curves to demonstrate the expansion of the ejecta in time (Figure \ref{fig:t_r}). The time for SNe Ia to reach maximum brightness  from the time of explosion ($t_{\rm max}$) ranges from $15$ to $22$ days according to \cite{miller20}. We use the upper value of 22 days to convert the observed phases into days since the explosion to avoid negative time values on the x axis of Fig. \ref{fig:t_r}.

We can see that the Ca II HVFs reach a distance of $\approx 3.0-4.0 \cdot 10^{15}$cm from the WD by $t_{\rm max}$, while the PVFs lag behind at around $\approx 2 \cdot 10^{15}$ cm. Following \cite{thomas04}, we calculate approximate masses for the HVF layer. For a C/O rich composition reminiscent of the unburned outer layers of an exploded WD, they estimate a minimum density of $\rho_{\rm min} \approx 2.2 \cdot 10^{-16}$ g cm$^{-3}$ that is needed for Ca II NIR3 lines to be visible. If we assume that HVFs arise from a spherically symmetric shell having a radius of $\sim 3$ - $4 \cdot 10^{15}$ cm, we get a minimum mass of this HVF-forming shell as $3.4 \cdot 10^{31}$ g or $\approx 0.017 M_{\odot}$.

\section{Discussion}
\label{sec:discussion}

Although HVFs are ubiquitous \citep{mazzali05b, maguire14}, their basic characteristics (strength, velocity, time to fade) vary from object to object, while collectively showing a connection to the stretch of the light-curve (\cite{childress14, maguire14, silverman15}, and Sec \ref{sec:x1_correlations}). 

A model for their formation must naturally explain
\begin{itemize}
    \item the strength and velocity variation from object to object, 
    \item the slowing and fading of HVFs in each event (i.e. the temporal evolution of HVFs),
    
    \item the velocity and strength differences between Ca II and Si II HVFs,
    \item the apparent sparsity of HVFs in HV events,
    \item the lack of HVFs in 91bg-like SNe Ia,
    \item the occasional appearance of HVFs in not just Si II and Ca II but also Fe II, O I and C II,
    \item the polarization of HVFs, especially of Ca II,
    \item the tendency of strong HVF objects to inhabit late-type (low mass, high sSFR) host galaxies,
    \item and most importantly the connection of the HVF velocity and strength to the light-curve stretch.
\end{itemize}

The emergence of HVFs is often attributed to one of the following three effects (or a combination thereof): an abundance enhancement (AE) of IMEs (intermediate mass elements) like Si and Ca in the outer parts of the main ejecta, a density enhancement (DE) outside the main ejecta, or ionization effects (IE).

An abundance enhancement of the observed elements (primarily Si and Ca) at high-velocity regions will inherently be tied to the explosion mechanism, or processes that occur shortly before the explosion. This could mean significant burning on the surface of the WD or in a He shell surrounding it prior to or during the explosion. While an abundance enhancement can occur according to multiple models, there is no natural way for these enhanced regions to become asymmetric or clumpy, in contradiction with the polarization measurements.

High abundances of Si could, for example, be reached by He-flashes on the surface of a CO WD that is accreting helium-rich material from the binary partner \citep{kato18}. The production of Ne, Mg and Si by He-shell flashes grows with increasing WD mass and accretion rate. This could result in a Si-rich layer by the time the WD reaches $M_{\rm WD}\approx 1.35~ M_{\odot}$, but for this mechanism to produce sufficient amounts of Ca, the WD has to be quite massive ($M_{\rm WD} > 1.38 ~M_{\odot}$). After explosion, these abundance-enhanced layers are thrown off, with Ca being in the farthest layer out. This would be one way to explain the velocity difference usually found between the Ca II and Si II HVFs. Since high masses are needed to reach the necessary abundances, this model of HVF production is only possible if all SNe Ia that show HVFs explode very close to the Chandrasekhar mass. A growing number of observations, however, suggest that a non-negligible fraction of Type Ia SNe (including those that display HVFs) explode at sub-Chandrasekhar masses \citep[e.g.][]{scalzo14, ktr20, bora24}.

A density enhancement would mean the accumulation of material at high velocities, and can either be attributed to the explosion mechanism itself (through asymmetric explosion, or ejection of high velocity material) or to interaction with circumstellar material.

Another possible mechanism for creating HVFs is ionization effects. HVFs of Ca II could be born even in a regular, spherically symmetric ejecta because of ionization changes at high velocity that lead to larger optical depths \citep{blondin13}. No abundance or density enhancement is necessary for this to occur, but such ionization effects only give rise to HVFs in the Ca II NIR triplet (and H\&K, but with different ionization properties), and not in the Si II line or other ions known to produce HVFs.

In this section, we examine the possible scenarios in detail and speculate about the possibility of them being the cause of HVFs.

\subsection{Interaction with binary partner}
Although HVFs do not appear in any published binary interaction models, SNe Ia are thought to emerge from binary systems, and so it is still reasonable to examine the possibility of HVFs being the product of interaction with the companion star. 

\cite{kasen10} examines interaction with a non-degenerate companion, and finds that depending on the viewing angle, only in around $10\%$ of cases can a strong UV/blue optical excess emission be detected as a signal of the interaction. On top of this, \cite{noebauer17} modeled the early light curves resulting from different explosion mechanisms, and found that the signal of such an interaction with a companion star is almost indistinguishable from the signal of a double-detonation explosion or interaction with close CSM. This means that the presence of an early bump in the LC is not necessarily a smoking gun for ejecta-companion interaction.

If the HVFs are somehow the product of such an interaction, it is safe to assume that similarly to the early bump in the LC, their detection would heavily depend on the viewing angle, thus, they would be more rare than it is currently observed (they are seen in $\approx 90\%$ of cases). 

Weak viewing angle effects in general could account for the variation of strength and velocity between objects, but it is unlikely they are connected to the collision between the supernova ejecta and the binary partner.

Even if HVFs are not formed from this interaction, could it somehow affect their evolution? We observe the HVFs to expand homologously, reaching a distance of $\approx3.5\cdot 10^{15}$cm by the time of maximum light. This distance is much larger than the size of a WD,  of any possible companion star, or the typical separation for a nova system. This implies that if there was an interaction with a companion star, it occured in the very first moments after explosion, and did not leave a mark on the homologous expansion of the HVF region and the photosphere. Considering everything, we believe this to mean that not only is companion interaction unlikely to be the source of HVFs, but even if its was, its effect on the expanding ejecta would be negligible.

\subsection{Interaction with CSM}
\label{sec:csm}
It has been shown \citep{gerardy04} that interaction of the ejecta with solar-abundance, low mass circumstellar material can create a dense, high-velocity shell which will then lead to HVFs of Ca II (and of weak O I). 

This CSM could be either $i)$ the accretion disk, $ii)$ a shell of material on top of the WD, $iii)$ the wind of the binary partner, or $iv)$ the result of mass loss from the binary system. In a spin-up/spin-down model, some CSM could be left around the primary WD, and provide material with which the ejecta can interact.

Whatever the nature of the CSM, it would most likely be asymmetrical and lead to viewing angle effects. The earliest observed HVFs have a velocity of $\approx 35,000$ km~s$^{-1}$ 1-2 days after explosion. If they are formed by interaction with CSM, it has to be located at a distance of $< 3 \cdot 10^{14}$ cm from the WD. This rules out the more distant options like CSM shells swept up by winds, or ISM surrounding the system. In order for us to observe these high velocity spectral lines in the earliest spectra, the CSM must by intrabinary, close to the WD, possibly the product of accretion.
A compact ($< 1R_{\odot}$) CSM shell would mean there is no extra light visible at early times in the LC, as the light produced by the interaction would already fade by the time the observations can be started.  

Specifically, in the case of SN~2011fe \cite{mulligan18} showed that the main SN ejecta alone cannot explain the observed Ca II NIR3 HVF absorption. Instead, the HVF evolution was better fit by a model of the ejecta interacting with a compact circumstellar shell of around $0.005~M_{\odot}$.

\cite{mulligan17} showed that the density profile of the shell would affect the line shape, and found that a reverse-sawtooth density profile describes the observations the best. They found that the mass of the shell affects the strength of the HVF absorption at the earliest phases, as well as the time of transition from absorption dominated by the shell material to absorption dominated by the (photospheric) ejecta material. According to their models, the time of transition occurs at later phases with increasing shell masses. We can use the $t_{\rm switch}$ parameter (the phase at which $R_{\rm HVF}$ reaches unity) from Section \ref{sec:gauss} as a proxy for this time of transition.

There does seems to be a tendency in our sample for objects with stronger Ca II NIR3 HVFs to reach $R_{\rm HVF}=1$ at later phases. If $t_{\rm switch}$ can truly be used to track the mass of the shell, this would mean that stronger Ca II HVFs are the result of more massive CSM shells. We find a positive correlation (with $p = 0.038$) between $t_{\rm switch}$ and the average Ca II NIR3 $R_{\rm HVF}$, but no significant relation between $t_{\rm switch}$ and $v_{\rm Si II}^{\rm tmax}$.

Blueshifted, time-varying Na I D absorption lines in supernova spectra have also been used as tracers of outflowing circumstellar material within, or surrounding the progenitor system, at distances of around $\approx 10^{17}$ cm \citep{sternberg11}. 

It seems that SNe Ia with Na I D absorption tend to have broader light curves, regardless of whether or not the absorption is red- or blueshifted \citep{maguire13}. 

We can also observe this somewhat in our own sample, which contains 12 supernovae with clear Na I D absorption. Out of the 12, we have the light curve stretch parameter for 8 of them, and 6 of those have SALT $x_1$ > 0.0, putting them in the slow-declining category. Our spectra however are not high-resolution enough to examine the Na I D lines in detail, and we can only gain information on whether they are present or not.

\cite{clark21} investigated if this blueshifted, time-varying Na I D absorption, which could be a probe of CSM at larger distances ($\approx 10^{17}$ cm), shows a connection with the appearance of Ca II HVFs, which -- if caused by CSM -- could probe nearby ($\approx 10^{14}$ cm) CSM. They found no connection between the two, suggesting that either one, or both of these phenomena does not have a CSM origin. Alternatively, it could be that CSM at these two distance regimes cannot be probed simultaneously. Additionally, while HVFs are seen in almost all SNe Ia, blueshifted Na I D absorption only appears in $23-48\%$ of events \citep{clark21}.

Apart from all this, even if the HVFs are not directly caused by a CSM shell, such a shell can still play a role in the formation of HVFs. Mixing of the ejecta with low mass H-rich CSM can lead to wider and deeper HVFs because of the increase in electron density that encourages recombination of Ca III to Ca II \citep{mazzali05a, li26}. Thorough mixing of the CSM and ejecta material in the high-velocity regime is needed for this effect to take place.

\subsection{Consequence of the explosion mechanism}
Another possible origin of the HVFs could be something intrinsic to the explosion mechanism itself. It is possible that during the explosion, high-velocity material is ejected shortly before the main ejecta, creating a region in which HVFs can form. Since we see both high-velocity and photospheric components at the same time, this high-velocity secondary photosphere is most likely aspherical and does not fully cover the primary photosphere. This spatial asphericity was recognized early on from polarization measurements of HVFs and is often described as a clumpy second photosphere \citep{kasen03, thomas04, tanaka06, maund13, ni23}. This clumpy material must however have a large enough covering factor to allow us to observe HVFs from nearly any viewing angle, or we wouldn't be able to observe HVFs as often as we currently do.

An inherently aspherical explosion would explain the geometry deduced from polarization measurements.
It would also lead to minor viewing angle effects that in theory could explain the diversity of HVFs across objects. However, because of the correlations discovered in large samples between the characteristics of the HVFs and light-curve width \citep{childress14, maguire14, silverman15}, host type \citep{pan15} and local star formation history \citep{meng19}, it seems that the variation of HVFs is also linked to a physical reason rather than just random orientation.

The crux of the problem then is \textit{how} high-velocity material can be created during the explosion, and how it can be connected to the other properties of the SN. To answer this, we take a closer look at some of the most prominent explosion models.

\subsubsection{Off-center delayed detonation}
Delayed detonation models of near-Chandrasekhar mass WDs with an asymmetric or off-center ignition naturally produce Si and Ca rich regions in the farthest layers of the ejecta. After the deflagration phase, an off-center detonation ignites the remaining material and creates IME-rich areas in certain directions \citep{hoeflich21}. These high-velocity, abundance enhanced clumps of material can then create a secondary photosphere that produces HVFs atop the primary photosphere \citep{ni23}.

Some extremely strong HVFs can also be achieved by combining the abundance enhancement of off-center delayed detonation models with ionization effects. \cite{li26} explain the particularly deep Ca II NIR3 HVFs of SN~2024gy by mixing of the ejecta with H-rich CSM, as described in Section \ref{sec:csm}.

One issue with this explanation is that there are more and more indications that a significant fraction of SNe Ia (at least $\sim 50$ percent) explode significantly below the Chandrasekhar mass \citep[see e.g.][and references therein]{bora24}, while delayed detonation requires $M_{\rm WD} \approx M_{\rm Ch}$. 
As one example, from late-time nebular spectroscopy, \cite{flörs20} calculated the abundance ratio of Ni to Fe for 58 SNe Ia and found that $\approx85\%$ of their sample agreed with the predictions of sub-Chandrasekhar explosions.

Additionally, the presence or strength of HVFs do not follow the observed distribution of ejecta masses between the $M_{\rm Ch}$ and sub-$M_{\rm Ch}$ regimes. According to \citet{silverman15}, $\sim 91$ percent of normal SNe Ia spectra taken at $-4$d or earlier contain HVF of Ca II, which is inconsistent with the low percentage of near-$M_{\rm Ch}$ SNe found by previous works, if delayed detonation is the root cause for HVFs.

Using nebular phase spectra, analysis of the morphology of emission lines provides a way of gaining information about the explosion physics. With high resolution JWST mid-infrared spectra, this method can be applied to the line profiles of different ions to differentiate between explosion models. Using this, \cite{kwok26} determined the most probable explosion mechanism for SN~2024gy and SN~2021aefx to be a Chandrasekhar-mass delayed detonation. Both of these events have strong HVFs in their early spectra.

\subsubsection{Double detonation of a He-shell}
In the double detonation scenario, a He-shell accumulates on the surface of the CO WD (which does not have to be $\approx M_{\rm Ch}$). When the shell reaches a critical mass, it detonates, which also ignites the underlying carbon-oxygen core, leading to explosion \citep{fink10, shen10}. The newest double detonation models \citep{townsley19, boos21, shen21} use a thinner He-shell ($M < 0.1  M_{\odot}$) compared to the first studies, and are able to reproduce the majority of normal SNe Ia light-curves and spectra. In this explosion mechanism, the mass of the WD ranges from sub-Chandrasekhar masses ($M_{\rm WD} \approx 0.8 M_{\odot}$) to $M_{\rm Ch}$, and this increase in mass leads to brighter, slow-declining events. Models predict an early UV/optical excess light from the shell \citep{piro25}, and even a detection of He I in the NIR \citep{collins23} as possible signatures of this explosion mechanism.

IMEs like Si and Ca would be synthesized in the ignited He shell and would then be ejected at high velocities during the explosion, leading to a possible source of HVFs. 
Such an extremely off-center detonation would probably cause some sort of asymmetry in the expanding ejecta, at least in the high-velocity shell. The frequently observed significant polarization of HVFs is in accord with this expectation. Alternatively, the high-velocity shell could be clumpy to let the underlying inner photosphere also be observable. A clumpy, high-velocity ejecta with a large covering factor would mean that HVFs are visible from most viewing angles, which is in agreement with their observed ubiquity.

\cite{mulligan19} found that interaction of the ejecta with a He-shell containing solar-abundance Ca produces no Ca II HVFs. This is due to the calcium being in the form of Ca III instead of Ca II because of the higher ionization potential (leading to lower electron density) of a helium substrate. In this scenario, it would require supersolar abundances of Ca to accurately reproduce the Ca II HVFs.
Note, however, that \citet{mulligan19} used only a spherical kinematic model for the shell ejection, not a self-consistent detonation model, thus, their model cannot constrain either the shell asymmetry or the chemical composition.

\cite{michaelis25} modeled the quiescent He-accretion on a single-degenerate CO WD, and produced a sub-Chandrasekhar double detonation that agrees with the observables of normal SNe Ia. They used a thick He-shell that in previous studies did not reproduce normal SNe Ia, but they demonstrate that despite prior results in the literature, their thicker-shell model could also lead to normal SNe Ia.
They also mention the possibility of Ca II HVFs forming in the outer parts of the IME mantle, or in the high-velocity ashes of the He-shell. Their models only reach up to $22000$ km~s$^{-1}$, which is low for observed HVFs of Ca II. The angle-averaged mass fraction of Ca and Fe, both elements that show HVFs, display a peak in the high-velocity outer edge.

\cite{terwel26} compared TARDIS spectral modeling based on different explosion models to fit the observations of SN~2024gy. Most model spectra fit the observations reasonably well, but only the double detonation model was able to reproduce the Ca II HVFs observed in SN~2024gy. Other evidence in their paper (for example the stable Ni/Fe ratio) points to a delayed detonation scenario, so a definitive conclusion about the explosion mechanism could not be made, but the appearance of HVFs in the double detonation models shows promise. Additionally, similarly to us, they find that an HVF component is also needed to fit the early Si II absorptions, while \cite{li26} did not.

Recently, the discovery of a double shell morphology in supernova remnant SNR 0509–67.5 was also proposed as evidence for double detonation \citep{das25}. Here, the outer Ca shell could be the remnant of the He-shell and the source of HVFs.

In the context of a double detonation model, the range of HVF strengths and velocities could be directly related to the mass of the WD and/or the shell. There is a positive correlation between SALT $x_1$ and the mass of the ejecta ($M_{\rm ej}$) \citep{scalzo19, bora24}, and since the HVF properties (strength, velocity) seem to correlate with $x_1$, it is possible this connection extends to the mass of the ejecta as well.
In our sample, we have information about $M_{\rm ej}$ in only 7 cases \citep[from][]{ktr20,bora24}, which is not enough to draw definitive conclusions. We do see a trend of SNe Ia with higher $M_{\rm ej}$ having higher average HVF velocities and strengths, but, with p-values above $0.05$, both these correlations lie above the threshold for statistical significance, and this avenue warrants further investigation with a larger sample.

For studying HVFs, early-phase spectra are needed, because HVFs fade shortly after maximum. For determining the mass of the ejecta however, spectra from around the time of maximum is used along with multi-color light-curves extending from pre-maximum phases to 40+ days beyond maximum. So in order to investigate both the evolution of HVFs and the mass of the ejecta, almost constant spectroscopic and photometric observation is needed from the time of discovery to more than a month post-maximum. This is not possible for most SNe Ia, leading to a small sample size in our analysis.

Finally, current double detonation models assume thin He-shells that are around the mass of 0.02-0.1 $M_{\odot}$ \citep{gronow21, boos21}. This mass range is in agreement with our rough estimate for the minimum mass of a shell ($M \gtrsim 0.017 ~M_\odot$) needed to produce Ca II NIR3 features in the HVF region (Section \ref{sec:t_radius}).

\subsection{Consequence of progenitor and/or environment}
When we look at the properties of SNe Ia, it is important to remember that they do not explode in a (figurative) vacuum, but that they are part of a progenitor system, a local stellar environment and a host galaxy that most likely influenced their evolution from the birth of the precursor star all the way to the explosion.

Indeed, there are clues hinting at the possibility of SNe Ia being influenced by their immediate or wider environment, characterized by the morphology, mass, age, color, metallicity or specific star formation rate (sSFR) of their host galaxy. 

For one, it has been shown that the peak luminosity (and light-curve shape) is correlated with the morphological type of the host galaxy. Late-type, bluer galaxies hosting younger stellar populations tend to produce brighter SNe Ia  when compared to early-type, red hosts \citep{hamuy96, pruzhinskaya20}. This connection between the properties of individual SNe Ia  and their hosts is also apparent in the so-called ``mass-step": SNe Ia in more massive (presumably metal-rich), low sSFR galaxies tend to be fainter \citep{kelly10, sullivan10}. Because metallicity increases with the age and mass of a galaxy, it is reasonable to assume that metallicity is one of the key parameters driving the variation across events exploding in different host galaxies.

Another clue is connected to the NV/HV classification of \cite{wang09}.
HV events tend to prefer the inner (presumably metal-rich) regions of their hosts, while NVs appear in the outer regions \citep{wang13}. Moreover, HV SNe Ia prefer more massive, metal-rich host galaxies, while NVs appear in all types of hosts \citep{pan15, pan20, pan24}. 
\cite{pan20} argue that this could be evidence for two populations of SNe Ia, one of which is sensitive to metallicity, leading to the HV supernovae. 

Since it appears that HV events have higher HVF velocities (Fig. \ref{fig:wang_HVF}.), but a smaller range of $R_{\rm HVF}$ values compared to NV events (Fig. \ref{fig:si_RCa}.), perhaps whatever is the reason for the NV/HV separation may also be connected to the formation of HVFs.

Events with blueshifted, time-varying Na I D absorption in their spectra (often used as a sign of CSM, see Section \ref{sec:csm}.) tend to be brighter and most often appear in late-type host galaxies as well \citep{maguire13}. This is to be expected, because these types of galaxies are still in the process of active star formation and tend to host a younger population of stars, more prone to be surrounded by nearby material.

Some connection with the host galaxy can be found regarding the SN Ia subclasses as well. When we look at the host properties of faint SNe Ia, they tend to appear in massive red galaxies that typically host older stellar populations \citep{gonzalezgaitan11}. \cite{hakobyan20} found that spectroscopically identified dimmer 91bg-likes occur more often in early-type galaxies, while brighter 91Ts prefer late-type hosts.

Overall, it seems that the nature of the host galaxy (age of the stellar population, mass, metallicity) has an influence on the most basic properties of the explosion (peak brightness, color, ejecta velocity). Based on this, it should be tested if this influence also extends to the properties of HVFs themselves. 
Indeed, SNe Ia with strong HVFs around the time of maximum brightness have been connected to low mass, later type hosts or to galaxies with high sSFR \citep{pan15, meng19}. On the other end of the spectrum, HVFs are notably absent, or exceedingly rare in 91bg-like events \citep{silverman15}, which occur more often in early-type galaxies. This could mean that for some reason, HVFs find it difficult to form in the high-metallicity, high-mass, low-sSFR environments that usually host fainter SNe Ia like 91bgs.

One explanation for such a correlation between HVF properties and the host galaxy could be the presence of circumstellar material in the system. In older (elliptical) galaxies, the high-mass stars prone to mass loss by stellar winds or Roche lobe overflow are already gone, leaving behind an old population of lower mass stars. The ISM is eventually depleted and star formation halts. So in older SNe Ia progenitor systems, CSM could be less common than in younger ones, because a low-mass companion star is not as prone to mass-loss as a more massive star would be.
In this case, if we presume that HVFs form by interaction of the ejecta with close CSM, younger systems (more common in late-type hosts) are more likely to have CSM around the WD, which could then lead to HVFs. Inversely, we would see almost no HVFs in SNe that explode in older galaxies, hosting older progenitor systems.

However, the connection between host galaxy age and the prevalence of CSM in SNe Ia progenitor binaries is not so trivial. As discussed in Section \ref{sec:csm}, for HVFs formed by CSM interaction to be visible in the earliest spectra, the CSM has to be very close to the WD. Mass loss swept up by stellar winds in the case of younger systems would be too far away to be a probable cause of HVFs. Closer forms of CSM, accretion disks or shells on the surface of a WD, could still form in binary systems with low-mass companions hosted in elliptical galaxies. In short, the type of CSM that is needed for the formation of HVFs is not dependent on the age of the companion star or host galaxy. Additionally, early type galaxies do still contain some interstellar material, including gas and dust \citep[see e.g.][and references therein]{simonian17}. Considering all of this, we find it unlikely that the connection between galaxy morphology and the HVF properties of SNe Ia are due to the age-related presence of CSM or ISM.

Another option is that late type host galaxies with active star formation have a lower-metallicity population of progenitor stars that evolve into CO WDs than those in early type galaxies. This leads to the conclusion that the age of the host galaxy -- through the metallicity of the progenitor star -- influences the nature of the resulting CO WD all the way to the point of explosion, including the amount of synthesized $^{56}$Ni \citep{timmes03, howell09}. According to the theoretical framework of \cite{timmes03}, higher metallicity progenitor stars evolve into WDs with more neutron-rich elements, leading to less synthetized $^{56}$Ni compared to stable burning products, producing less luminous SNe Ia. 
This seems to be at odds with the observation that more luminous supernovae explode in younger, higher metallicity environments, and the reason for this discrepency is still not clear.

Regardless of whether the correlation between progenitor metallicity and nickel mass were positive or negative, since the nickel mass is the primary driver of SNe Ia luminosity (for which SALT $x_1$ is a proxy), and HVF properties (velocity, strength) are correlated with SALT $x_1$, it is possible that the progenitor metallicity does not only affect the production of nickel, but also the formation of HVFs.

It has been shown before \citep{mannucci06, brandt10} that the delay time distribution (DTD) of SNe Ia can best be described by a bimodal function, consisting of a ``prompt" and a ``delayed" population of progenitors. The prompt channel leads to explosion within a few hundred Myrs of star formation ($M_{\rm ZAMS} = 5.5 - 8 \text{ M}_{\odot}$), while the delayed channel ($M_{\rm ZAMS} = 3- 5.5 \text{ M}_{\odot}$) has a time scale of a few Gyrs (agreeing with the theoretical prediction of the DD channel \citep{maoz12}). The prompt supernovae are typically high-stretch (bright) in late-type host galaxies, while the delayed population is made up of fainter events in early-type hosts \citep{sullivan06, brandt10, wojtak23}. This suggests that the SN Ia formation rate is connected to the mass and the star-formation history of the host galaxy, with higher SFR galaxies hosting more (and brighter) SNe Ia per unit stellar mass per year \citep{sullivan06}.

In this framework, HVF-strong events mostly appear in the prompt population, while HVF-weak SNe prefer the delayed population of early-type hosts.

\begin{figure}[ht]
    \centering
    \includegraphics[width=0.95\linewidth]{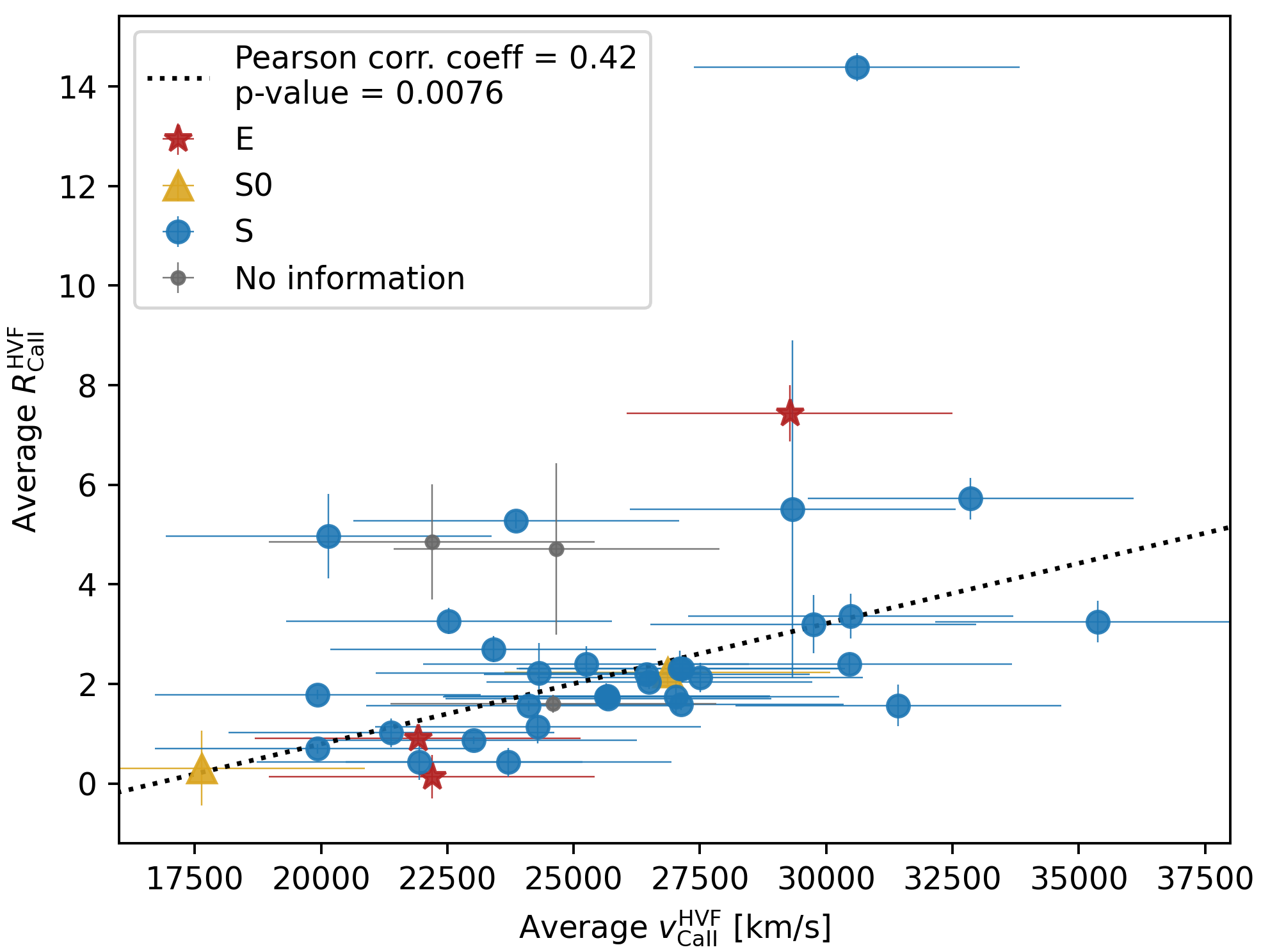}
    \caption{Average Ca II NIR3 HVF strength of each object compared to their average (SYNOW modeled) velocity. Blue dots mark the events with spiral host galaxies, yellow triangles show those with lenticular hosts and red stars present the elliptical hosts. The host galaxies without available morphological data are marked with small grey dots. The object with an elliptical host and strong, fast HVF is SN~2012hj. An overall positive correlation between the average strength and velocity of HVFs can be seen in the figure (dotted black line).}
    \label{fig:R_vel}
\end{figure}

Our sample cannot be used for precise analysis of host galaxy mass or SFR, so we used the morphological types of the host galaxies to generally characterize them. The morphologies were obtained from the SIMBAD database \citep{wenger00} and HYPERLEDA catalogue \citep{paturel03} when available. In the cases where a published morphological class was not available for the galaxy, we determined the morphology by visual inspection of the PanSTARRS DR1 color images. 
A rudimentary representation of the morphological type in Figure \ref{fig:R_vel} displays the expected outcome. Most of the host galaxies in our sample were spiral galaxies, with 3 elliptical and 5 lenticular hosts. When we look at the hosts in the context of the average (SYNOW modeled) Ca II HVF velocity and strength of a supernova, we see that most events in early-type galaxies have weak, slow HVFs. SN~2012hj is a notable exception, as it does not follow the expected trend. Its host is an elliptical galaxy, and so based on our assumptions, we would expect it to have weaker, lower velocity HVFs than it does. However, this SN exploded quite far away from the core of the host, in the less metal-rich halo region, which could explain why it is similar in characteristics to those exploding in lower-metallicity spiral galaxies.

Throughout this section, we often used the global parameters of the host galaxy to attempt to describe the local environment of the exploding WD. This is not ideal, because parameters like metallicity or presence of ISM are not constant within a galaxy. Local measurements of the stellar population near the SN locations would be preferable in order to draw robust conclusions, but this was not the goal of our current study.

When we consider all the previous studies conducted on the connection of SNe Ia properties (including those of the HVFs) to the host galaxy, we can construct Table \ref{tab:host_type}, similarly to Table 5 in \cite{maguire13}. Whether or not Table \ref{tab:host_type} represents two distinct families of SNe Ia, two ends of a continuum, or some other unexplained effect, is well beyond the scope of this paper.

\begin{table*}[ht]
    \centering
    \small
    
    \begin{tabular}{cccc}
    \hline
     \hline
         & Early type host  &   Late type host &    Reference\\
         & (high $M_{\rm stellar}$, low sSFR)   & (low $M_{\rm stellar}$, high sSFR)    &   \\
     \hline
     DTD & Delayed   &   Prompt  &    g, l, a, m  \\
     Peak luminosity    &   Faint    & Bright     &   e\\
     Wang-type   &   More HVs     &   Less HVs &  i, j, k\\
     HVF strength   &   Weak    &   Strong  &  i, h\\
     HVF velocity &     Slow    &   Fast    &   Fig \ref{fig:R_vel}. \\
     Na I D absorption  &   None, or Redshifted &   Blueshifted &   f, c\\
     91bg-like &   Common  &   Not common  &  d, b\\
     \hline
    \end{tabular}
    \caption{Generalized differences in Type Ia supernova properties (including properties of HVFs) based on their host galaxy. \\
    a: \cite{brandt10}, b: \cite{chakraborty24}, c: \cite{clark21}, d: \cite{hakobyan20}, e: \cite{hamuy96}, f: \cite{maguire13}, g: \cite{mannucci06}, h: \cite{meng19}, i: \cite{pan15}, j: \cite{pan20}, k: \cite{pan24}, l: \cite{sullivan06}, m: \cite{wojtak23}.}
    \label{tab:host_type}
\end{table*}

\section{Conclusions}
\label{sec:conclusions}

We explored the presence and evolution of HVFs of Ca II and Si II in a sample containing 181 optical spectra of 56 Type Ia SNe. For studying the line profiles, we applied both Gaussian fitting (similar to most previous studies) as well as spectrum synthesis (using SYNOW). We found that Gaussian fitting usually underestimates the velocity of the Ca II HVF, and derived a linear function that can be used to correct the Gaussian velocities to the more appropriate model velocities (Eq. \ref{eq:vs_vg}.). 

Regarding the connection between the properties of HVFs and other observables of SNe Ia, we found that, in accord with several previous studies, brighter, slow-declining SNe Ia have faster (Fig \ref{fig:x1_mosaic}), stronger HVFs (Fig \ref{fig:x1_R}) and higher velocity separation from their photospheric component (Fig \ref{fig:x1_gap}). Most of our sample SNe belonged to the Wang NV (normal velocity) type, only a few of them could be assigned to the Wang HV (high-velocity) type. The strength of HVFs in NV objects showed an anti-correlation with $v_{\rm SiII}^{t_{\rm max}}$ (Fig \ref{fig:si_RCa}). This anti-correlation between HVF strength and photospheric velocity seems to only apply to the NV objects in our sample, but it is difficult to draw conclusions because of our low number of HV events. We also find that HV events show higher Ca II HVF velocities than NVs (Fig \ref{fig:wang_HVF}). This might be an indication that HVFs in HV and NV supernovae behave differently.

We find that the stronger an HVF (of Ca II NIR3) is, the longer it takes for it to fade. In the CSM interaction models of \cite{mulligan17}, this time of fading (characterized by $t_{\rm switch}$ in this paper) is longer with larger CSM masses. We also see a tentative ($p = 0.07$) connection between the ejected mass of the supernova, and the average strength and velocity of the Ca II NIR3 HVF, but due to low sample size, we cannot draw definitive conclusions.

Comparing the inferred properties of HVFs with the morphological type of the host galaxies, our data suggest that SNe in early-type (elliptical, lenticular) galaxies exhibit weaker and slower HVFs compared to SNe in late-type (spiral) galaxies (Fig \ref{fig:R_vel}). 

When putting these observational findings together in context with the theoretical predictions, we conclude that the physics of HVFs is a complex phenomenon, and their root cause remains a mystery. 

By studying the evolution of HVF and PVF properties over time, it's most likely that their fading is due to changes in the temperature and density of the ejecta caused by the homologous expansion, and not because they eventually slow and blend with their photospheric counterparts. The velocity and optical depth separation of the Ca II NIR3 and Si II $\lambda6355$ \AA\ HVFs is also likely due to temperature differences within the ejecta, and not necessarily a sign of composition stratification.

As for their origin, we exclude interaction with a companion star or ISM as the source of HVFs because of strong viewing angle dependence and excessive distances, respectively. In order for external material to be the cause of HVFs, it has to be at a distance of $< 3\cdot 10^{14}$ cm from the WD at the time of explosion. With this distance scale, a compact CSM shell is still a likely candidate for the origin of HVFs, but the lack of connection with more distant CSM probed by the blueshifted Na I D absorption \citep{clark21} needs to be explained.

A more promising model to explain the presence of HVFs seems to be the double detonation explosion that can produce the necessary high-velocity shell, asymmetries, the range of observed parameters (perhaps influenced by the WD mass) and the observed double-shell morphology in some SNe Ia remnants \citep{das25}.

The HVFs themselves are most likely the result of multiple physical effects: a combination of abundance and density enhancements caused by the explosion mechanism itself, perhaps with some influence of the external global environment via the metallicity of the progenitor star. Density enhancements could result from either the shell of a double detonation, asymmetric blobs from an off-center delayed detonation, or clumps created naturally by Rayleigh-Taylor instabilities in the outer layers by the detonation.  On top of all this, viewing angle effects cannot be ruled out as the reason for \textit{some} variety in the observed HVFs, but they are unlikely to cause the majority of the variation across explosions.

Unfortunately, despite decades-long observational efforts, like our program, the observed sample is still too poor to draw any better statistical conclusion. 
The big picture is surely much more complex, and intimately tied to our questions about the progenitor systems and explosion mechanisms of SNe Ia.

We emphasize the need for non-LTE 3D radiative transfer simulations based on multiple different explosion mechanisms. In the almost 3 decades since HVFs have been discovered, a multitude of observations have been collected, analyzed, and published, but so far there has been a dearth of models to compare to the observations, mostly due to the computational price of such simulations.

\begin{acknowledgments}

We thank the anonymous referee for their helpful comments.

This research has been supported by NKFIH-OTKA Grant K142534. The SN research group at Konkoly received funding from the GINOP 2.3.2-15-2016-00033 grant from the Government of Hungary, funded by the European Union.  

The Hobby–Eberly Telescope (HET) is a joint project of the University of Texas at Austin, the Pennsylvania State University, Ludwig-Maximilians-Universitat Munchen, and Georg-August-Universitat Gottingen. 
The HET is named in honor of its principal benefactors, William P. Hobby and Robert E. Eberly. The Low Resolution Spectrograph 2 (LRS2) was developed and funded by the University of Texas at Austin McDonald Observatory and Department of Astronomy, and by Pennsylvania State University. We thank the Leibniz-Institut fur Astrophysik Potsdam (AIP) and the Institut fur Astrophysik Goettingen (IAG) for their contributions to the construction of the integral field units. The authors are grateful to the HET Resident Astronomers and staff members at McDonald Observatory and Las Cumbres Observatory for their excellent work.
This research has made use of the SIMBAD database, CDS, Strasbourg Astronomical Observatory, France.

\end{acknowledgments}

\begin{contribution}

ZB made most of the spectroscopic modeling and analysis, wrote the manuscript and served as a corresponding author.
JV was responsible for managing the observations and data reductions for most of the HET spectra. He also provided ideas for the modeling and analysis of the spectra.
JCW initiated the project and served as the PI of the observing proposals during almost two decades. He also contributed to writing the paper.
RKT and GHM contributed to the reduction of some HET spectra, and provided insights to the interpretations of the results. The authors express their thanks to J. M. Silverman for his contribution to the HET observations.

\end{contribution}

\facilities{HET}

\bibliographystyle{aasjournal}

\section{Appendix}
\label{sec:appendix}

\begin{table*}[h]
    \centering
    \small
    
    \begin{tabular}{lr}
     \hline
       \hline
       SN   &  Source(s) \\
       \hline

    SN~2011fe  & \cite{pereira13}  \\
    
    SN~2012fr  & \cite{childress13}  \\

    SN~2013dy  & deimos \citep{zheng13}, \\
        &       DAO Plaskett \citep{balam16} \\
    
    SN~2017cbv  & LCO \citep{hosseinzadeh17} \\
     
    SN~2017hjy  & Ekar AFOSC \citep{benetti17} \\
    SN~2017ln  & EPESSTO \citep{taddia17} \\

    SN~2018gv  & Keck 1 \citep{siebert18} \\

    SN~2018ilu & ZTF \citep{fremling18}, \\
                & EPESSTO \citep{berton18} \\
    SN~2018oh  & \cite{leadbeater18}, \\
                & Lijiang YFOSC \citep{zhang18}  \\
      
    SN~2020aabz  & EPESSTO \citep{ihanec20}  \\

    SN~2021J   & ZTF \citep{dahiwale21}, \\
            & ePESSTO+ \citep{gillanders21} \\
    SN~2021gtp  & Lick \citep{hung21} \\
   
    SN~2021hiz  & SOAR \citep{siebert21} \\
    SN~2022erw  & EPESSTO \citep{moore22} \\
       
    SN~2022hkc  & \cite{balcon22}  \\
   
    SN~2022yhl  & Keck1 \citep{tinyanont22} \\
     
    SN~2023alb  & Keck1 \citep{davis23} \\
        
    SN~2023bee  & Lijiang YFOSC \citep{zhai23bee}, \\
            &   Global SN Project \citep{hosseinzadeh23tns} \\
     
    SN~2023eoc  & Lijiang YFOSC \citep{zhai23eoc} \\
      
    SN~2023ktw  & SCAT \citep{hinkle23} \\
       
    SN~2023pbe   & GOTO \citep{godson23}, \\
                & ZTF \citep{fremling23} \\
      
    SN~2023tky   & SCAT \citep{woods23} \\
       
    SN~2023vyl   & ZTF \citep{sollerman23} \\
       
    SN~2023xtf   & Ekar AFOSC \citep{schuldt23} \\
   
    SN~2023xuz   & ePESSTO+ \citep{kopsacheili23} \\
    SN~2023zvn   & ePESSTO+ \citep{pessi23} \\

    SN~2024gy  & Global SN Project \citep{newsome24}, \\
                & \cite{leadbeater24}, \\
                & \cite{borland24}, \\
                & \cite{balcon24} \\
  
    SN~2024iei   & NOT ALFOSC \citep{meissner24} \\
     SN~2024xyn  & Lijiang LiONS \citep{li24}, \\
                & ZTF \citep{wise24}  \\
    
    \hline
    \end{tabular}
    \caption{Table of all the additional spectral data used in our work in the case of each supernova. More detailed table can be found in the \dataset[data repository]{https://doi.org/10.5281/zenodo.22079031} accompanying this paper.}
    \label{tab:spectral_data2}
\end{table*}

\end{document}